%% file: paper.tex
\documentclass[a4paper,11pt]{article}
\pdfoutput=1 

\usepackage{sty/jinstpub} 

\usepackage{xcolor}

\definecolor{forestgreen}{rgb}{0.0, 0.27, 0.13}
\definecolor{darkpastelgreen}{rgb}{0.01, 0.75, 0.24}

\usepackage{lineno}

\usepackage{siunitx}
\DeclareSIUnit\atm{atm}

\usepackage[noabbrev]{cleveref}

\crefname{equation}{Equation}{Equations}
\crefname{reference}{Ref.}{Refs.}
\crefname{chapter}{Chapter}{Chapters}
\crefname{appendix}{Appendix}{Appendices}
\crefname{section}{Section}{Sections}
\crefname{figure}{Figure}{Figures}
\crefname{table}{Table}{Tables}

\usepackage{subfigure}
\usepackage[version=4]{mhchem}

\usepackage{sty/relsec}
\usepackage{sty/siunitx-hep}

\graphicspath{{./img/}{./diag/}{./plots/}}
\def\graphicsmode{_adj_rast}%

\usepackage{rotating}
\usepackage{multirow}

\usepackage{longtable}
\usepackage{float}

\newwrite\graphics
\immediate\openout\graphics=\jobname.graphics%
\let\oincludegraphics\includegraphics
\renewcommand{\includegraphics}[2][]{
  \immediate\write\graphics{#2}
  \oincludegraphics[#1]{#2}}

\title{Measurement of the fractional radiation length of a pixel module for the CMS Phase-2 upgrade via the multiple scattering of positrons}%

\collaboration{\includegraphics[height=17mm]{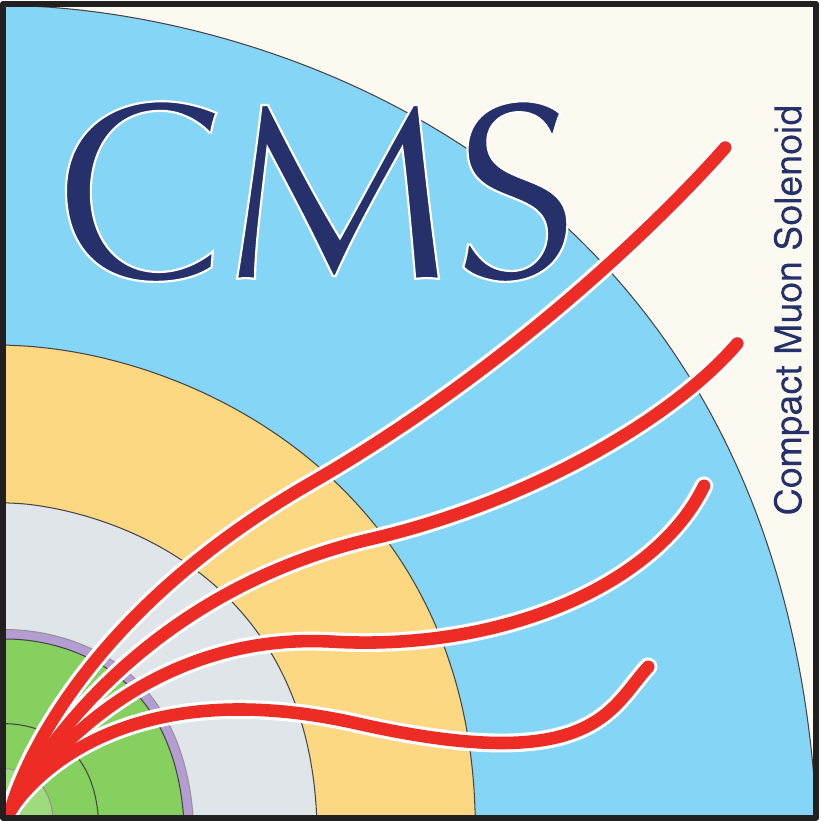}\\[6pt]%
   The Tracker Group of the CMS Collaboration}%

\emailAdd{simon.florian.koch@cern.ch}

\DeclareSIUnit{\Gbit}{Gbit}
\DeclareSIUnit{\Mbit}{Mbit}

\newcommand{\cmsAuthorMark}[1]{\hbox{\textsuperscript{\normalfont#1}}}

\newcommand{\Molex}{Molex\textsuperscript{\textregistered}~}%

\abstract{%

    High-luminosity particle collider experiments such as the ones planned at the High-Luminosity Large Hadron Collider require ever-greater vertexing precision of the tracking detectors, necessitating reductions in the material budget of the detectors.
    Traditionally, the fractional radiation length ($x/X_0$) of detectors is either estimated using known properties of the constituent materials, or measured in dedicated runs of the final detector. 
    In this paper, we present a method of direct measurement of the material budget of a CMS prototype module designed for the Phase-2 upgrade of the CMS detector using a 40--\SI{65}{MeV} positron beam. 
    A total of 630 million events were collected at the Paul Scherrer Institut PiE1 experimental area using a three-plane telescope consisting of the prototype module as the central plane, surrounded by two MALTA monolithic pixel detectors. 
    Fractional radiation lengths were extracted from scattering angle distributions using the Highland approximation for multiple scattering. 
    A statistical technique recovered runs suffering from trigger desynchronisation, and several corrections were introduced to compensate for local inefficiencies related to geometric and beam shape constraints.
    Two regions of the module were surveyed and yielded average $x/X_0$ values of $(0.72 \pm 0.05)\%$ and $(0.95 \pm 0.09)\%$, which are compatible with empirical estimates for these regions computed from known material properties of 0.753\% and 0.892\%, respectively. 
    Two types of higher-granularity maps of the fractional radiation length were produced, subdivided either into rectangular regions of uniform size, or polygonal-shaped regions of uniform material composition.
    The results bode well for the CMS Phase-2 upgrade modules, which will play a key role in the minimisation of the material of the upgraded detector.

}


\begin{document}
\maketitle
\flushbottom

\begin{asection}{Introduction}%
\label{s:introduction}

A move toward higher-luminosity particle accelerators necessitates continual improvement of silicon detector technologies, particularly in the areas of vertex resolution and radiation tolerance. As part of the CMS and ATLAS Phase-2 upgrades to meet the future operating requirements of the High-Luminosity LHC~\cite{cms2008,cms2023,cms_phase2_tdr_tracker,atlas2008}, the CERN RD53 Collaboration has developed readout chips (ROCs) based on a \SI{65}{\nm} CMOS process, with two main design iterations~\cite{rd53a,rd53b}. The RD53A architecture is shared between the CMS and ATLAS Collaborations, whilst the flavour of the RD53B chip design specific to CMS is referred to as the CMS Readout Chip (CROC). The CROC design allows for improvements on the resolution and readout rate of the current CMS Phase-1 pixel detector, reducing the sensor pixel size from $150 \times 100\,\si{\um^2}$ to $100 \times 25\,\si{\um^2}$ (corresponding to $50 \times 50\,\si{\um^2}$ readout cells), and increasing the maximal data rate per ROC from \SI{160}{Mb/s} to \SI{5.12}{Gb/s}.

    Within tracking detectors, uncertainties in vertex position are strongly influenced by the lever arm and material content of the innermost tracking layer, due to the higher impact of multiple scattering in layers closer to the interaction. As a result, substantial efforts are made to minimise the material budget of pixel-type detectors for current and future tracking detectors. Precise knowledge of the fractional radiation length contribution of module components is important both during the design phase for new detectors and after data-taking, as accurate spatially-resolved estimates or measurements are required in order to correctly simulate multiple scattering from the detector geometry in Monte Carlo simulations. 
    
    During design and development phases of detectors, the material budget is usually estimated from the material composition of the components in the design specification. The accuracy of such estimates is often limited by the knowledge of the composition of the individual components used, and such estimates are generally given as a global estimate for the material budget rather than a spatially-resolved map. Experimental confirmation of the theoretical estimates would both verify the integrity of current estimates, and improve knowledge of the detector geometry for implementation in simulations of secondary interactions in the detector.
    
    Verification of the material budget of CMS has previously been performed using alignment and calibration data~\cite{migliore2011}. In these measurements, photon conversions and nuclear interactions from track reconstruction are used to probe the $X_0$ (radiation length) and $\lambda_I$ (nuclear interaction length) distributions, respectively. Such measurements have previously been used to confirm that material in simulations generally match data to within $\sim$10--20\%, in particular via measurement of the energy loss of low transverse momentum electron and charged hadron tracks~\cite{khachatryan2015,sirunyan2018,cms-pas-trk-10-003}. Although important for the characterisation of the detector as a whole, these measurements are performed during or after commissioning of the detector and are hence too late to provide input to the detector R\&D process, and generally provide a relatively coarse spatial resolution. 

    In this paper we present a method to measure the fractional radiation length of the complex material stack in pixel modules via a measurement of multiple scattering with an \si{\MeV}-scale positron beam, achieving a fine-grained spatial resolution on the order of millimetres. The object of this study is a prototype quad module developed for the CMS Phase-2 pixel detector upgrade for the High-Luminosity LHC, and data from the module was utilised to provide the central point of the reconstructed deflection angles for the measurement. Studies of fractional radiation length via multiple scattering with \si{\GeV}-scale electron beams have been performed previously by groups from the Georg-August-University G\"ottingen and the German Electron-Synchrotron (DESY), but have relied on full telescopes for tracking rather than using data from an active subject, and have generally measured thicker or less complex subjects~\cite{stolzenberg2017,jansen2018,stolzenberg2019,arling2022}. Using a lower beam energy increases the mean scattering angle and reduces the impact of the telescope plane resolution, but in turn increases acceptance losses when imaging regions further from the centre of the subject.

    \begin{asection}{Multiple scattering and the Highland approximation}%
    \label{s:ms}
                
        The Bethe prescription of Moli\`ere scattering provides a robust approach for modelling multiple scattering due to its close agreement with experiment~\cite{bethe1952,lynch1990}, with an accuracy of 1\% or better shown by Hanson et al.~\cite{hanson1951}. The distribution of scattering angles is given as the sum of three analytic terms comprising a Gaussian core, a carry-over to the Rutherford formula at large angles (corresponding to large-angle single scatters), and a correction.    

        The analytic complexity of Moli\`ere's theory is often a barrier for its use in applications. Highland et al.~\cite{highland1975} fitted the Moli\`ere-Bethe-Hanson theory to derive an expression for the central Gaussian core of the distribution, and Lynch and Dahl's revised version~\cite{lynch1990} gives the root-mean-square (RMS) projected scattering angle of the central Gaussian as
        \begin{equation}
            \label{e:highland}
            \Theta := \theta^\text{RMS}_\text{plane} = \frac{\SI{13.6}{\mega\electronvolt}}{\beta c p} z \sqrt{x/X_0} \bigl (1 +0.038 \ln\left(x/X_0\right)\bigr).
        \end{equation}
        Here $\beta c$, $p$ and $z$ are the speed, momentum, and charge number of the scattering particle, respectively, and $x$ is the length of the particle trajectory through a material with radiation length $X_0$ (in \si{\cm}). Hence, the dimensionless quantity $x/X_0$ can be used to represent the fractional radiation length of the scatterer without referencing a specific trajectory. \Cref{e:highland} may be analytically inverted by substitution of the principal branch of the Lambert $W$ function~\cite{Corless1996} to give
        \begin{equation}
            \begin{aligned}
                \frac{x}{X_0} &= \exp \left( 2 \cdot W_0\left( w \right)  - \frac{1}{0.038} \right), \\
                w             &= \frac{1}{2 \cdot 0.038} \, \exp \left(\frac{1}{2 \cdot 0.038} \right) \, \frac{\beta c p}{\SI{13.6}{\mega\electronvolt} \cdot z} \, \Theta.
            \end{aligned}
        \end{equation}

    \end{asection}

    \begin{figure}[t]%
        \centering%
        \includegraphics[width=.7\textwidth]{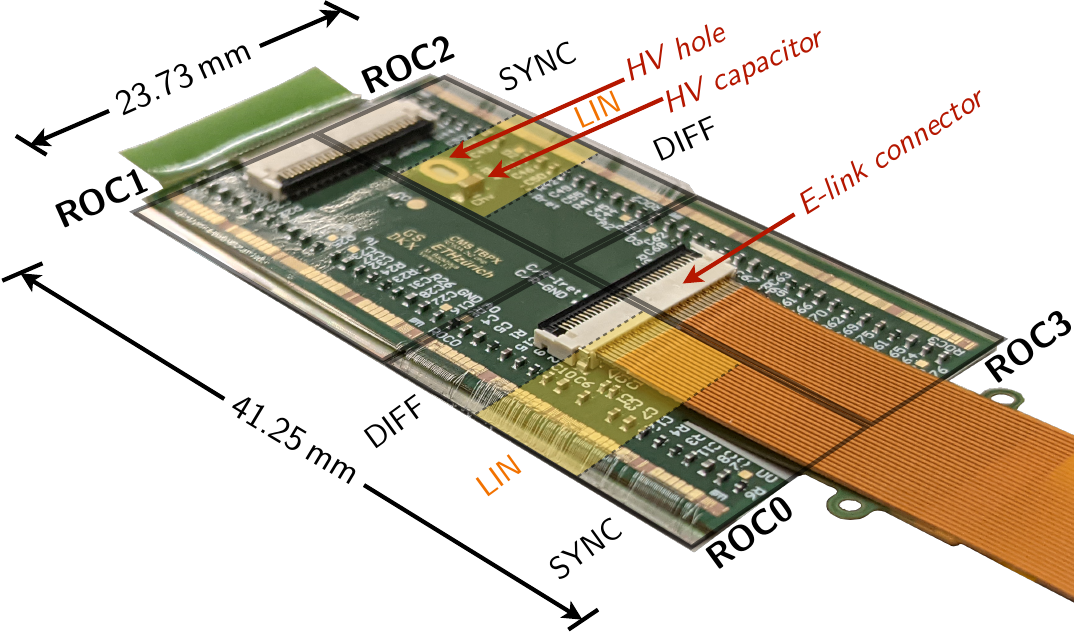}%
        \caption{\label{f:module_image}Photograph of the prototype pixel module used for this measurement. The linear front-end regions of ROC0 and ROC2 included in this measurement are shaded in yellow, and key features within these regions are indicated.}%
    \end{figure}

    \begin{asection}{The RD53A prototype 2\texorpdfstring{$\times$}{×}2 chips module}%
        \label{s:rd53a_quad}

        Shown in \cref{f:module_image}, the prototype pixel module studied in this measurement contains 2$\times$2 RD53A chips and was developed and qualified at ETH Zurich as an early prototype for layers~3 and~4 of the barrel of the CMS Phase-2 pixel detector design~\cite{backhaus2019,perovic2020}. The final upgraded pixel detector will consist of four barrel layers, eight small double-discs per side, and four larger double-discs per side. The layout will increase the forward coverage from $|\eta| \sim 2.6$ to $|\eta| \sim 4$~\cite{cms_phase2_tdr_tracker}. The total active surface area will be $\sim 4.9\, \si{\m^2}$, holding roughly $4000$ modules. 

        The prototype module design consists of four RD53A ROCs bump bonded to a planar sensor which is subsequently glued to a high-density interconnect (HDI) flexible printed circuit board (PCB). 
        The $192\times400$ pixel matrix of the RD53A chip includes three regions corresponding to different prototype discriminator circuits for evaluation: the linear, differential and synchronous analogue front-ends.
        The HDI provides data and power interconnects, and hosts passive components. Wirebonds form the connection from the HDI to the ROCs, and from the HDI to the high voltage (HV) bias pad on the sensor through a rounded hole in the HDI\@. Power is provided via a power pigtail integrated into the HDI, and a zero-insertion-force (ZIF) connector near the centre of the HDI accepts a data link. A second ZIF connector opposite the pigtail provides power output, allowing a row of modules on a ladder to be powered in a single serial powering chain, reducing power loss and cabling requirements. 

        Readout of RD53A ROCs is trigger-based, with a time resolution of \SI{25}{\ns} imposed by the bunch-crossing clock, as designed for compatibility with the LHC bunch-crossing interval. Data are digitised and packaged at the level of the readout chip, with a time-over-threshold value proportional to the total charge collected provided for each hit pixel when a trigger is received. The clocks within each ROC are derived with respect to an external reference bunch crossing clock, providing synchronisation for all ROCs on a prototype module.

        The measurements discussed in this paper were performed on a prototype module, constructed with a sensor produced by the FBK foundry in Trento with $\SI{50}{\um} \times \SI{50}{\um}$ pixels. The sensor has a total thickness of \SI{200}{\um}, including an active region of depth \SI{150}{\um} which is depleted during operation. The remaining depth is inactive handle wafer residual from the production process. The sensor was glued to a v3 tracker barrel pixel (TBPX) quad-module HDI~\cite{backhaus2018hdi}, and module rails composed of an AlN ceramic compound were attached to the ROC-side of the assembly. The material choice minimises the contribution of the rails to the material budget whilst providing thermal conduction and electrical isolation. 
    
        Due to cooling limitations in the testbeam setup, only two of the four ROCs on the prototype module were wirebonded to the HDI in order to halve the current consumption. Since the readout system for RD53A ROCs did not support simultaneous readout of multiple front-ends at the time of measurement, only the linear front-end was enabled for readout. ROC0 and ROC2 were selected for particular features of interest within the linear front-end region, as shown in \cref{f:module_image}. Following the evaluation of all three prototype front-ends, the Tracker Group of the CMS Collaboration has chosen the linear front-end for the CROC chip design to be used in the final Phase-2 detector~\cite{cmstracker2021}.

    \end{asection}

    \begin{asection}{Estimated material budget}%
        \label{s:estimate}

        \begin{table}[tbp]
            \caption{\label{t:estimate}%
                Material budget estimates for the components of RD53A TBPX FBK prototype quad modules~\cite{ristic2020}. Estimates for SMD (surface-mount device) components are not included in the intermediate totals with/without rails, since these are unevenly distributed and individually resolvable in the results presented. Separate estimates are given for the ROC0 LIN and ROC2 LIN regions from the HDI layout. Contributions with non-complete coverage (Cov.) are ``smeared'' across the relevant region as though they had 100\% coverage whilst retaining their volume. \\\textsuperscript{$\dagger$}Heights quoted for SMD components are averaged among all components in the region.%
            }%
            \begin{center}%
                \begin{tabular}{|l||ccccc|c|}
                    \hline
                    Region & Material & Cov. (\%) & $X_0$ (\si{\g.\cm^{-2}}) & $\rho$ (\si{g.cm^{-3}}) & $x$ (\si{\um}) & $x/X_0$ (\%) \\
                    \hline\hline
                    \bf Assembly &\bf - & \bf - & \bf - &\bf - &\bf 352.35 &\bf 0.394 \\
                        \quad ROC & Si & 100 & 21.82 & 2.33 & 150 & 0.160 \\
                        \quad Sensor & Si & 100 & 21.82 & 2.33 & 200 & 0.214 \\
                        \quad Bump bonds & Sn & 100 & 8.82 & 7.31 & 2.35 & 0.020 \\
                    \bf HDI (no SMDs) &\bf - &\bf - &\bf - &\bf - &\bf 160 &\bf 0.160 \\
                        \quad Traces & Cu & 76.5 & 12.86 & 8.96 & 23 & 0.123 \\
                        \quad Polyimide (PI) & PI & 100 & 40.58 & 1.42 & 24 & 0.008 \\
                        \quad Glues & Acrylic & 100 & 42.6 & 1.00 & 75 & 0.018 \\
                        \quad Coverlay & PI & 100 & 40.58 & 1.42 & 13 & 0.005 \\
                        \quad HDI glue & Acrylic & 100 & 42.6 & 1.00 & 25 & 0.006 \\
                    \hline
                        \bf Regions w/o rails &\bf - &\bf - &\bf - &\bf - &\bf 389.35 &\bf 0.553 \\
                    \hline
                    \hline
                        \bf Average module rails &\bf - &\bf 47 &\bf - &\bf - &\bf 275 &\bf 0.142 \\
                        \quad Ceramic rails & AlN & - & 27.5 & 3.26 & 250 & 0.296 \\
                        \quad Rail glue & Acrylic & - & 42.6 & 1.00 & 25 & 0.006 \\
                    \hline
                        \bf Regions w/ rails  &\bf - &\bf - &\bf - &\bf - &\bf 664.35 &\bf 0.856 \\
                    \hline
                        \bf Average (no SMDs)  &\bf - &\bf - &\bf - &\bf - &\bf - &\bf 0.696 \\
                    \hline
                    \hline
                        \bf SMDs  &\bf - &\bf - &\bf - &\bf - &\bf - &\bf 0.129 \\
                        \quad 87 SMDs 0402 & Ceramic & 2.82 & 11.16 & 2.6 & 445\textsuperscript{$\dagger$} & 0.052 \\
                        \quad 1 SMD 0603 & Ceramic & 0.083 & 11.16 & 2.6 & 800 & 0.003 \\
                        \quad 2 Connectors & Various & 9.5 & - & - & 1150 & 0.054 \\
                        \quad Flex cable & Various & 25.8 & - & - & 68 & 0.020 \\
                    \hline
                        \bf Average (w/ SMDs)  &\bf - &\bf - &\bf - &\bf - &\bf - &\bf 0.825 \\
                    \hline
                    \hline
                        \bf ROC0 LIN average  &\bf - &\bf - &\bf - &\bf - &\bf - &\bf 0.892 \\
                        \quad 7 SMDs 0402 & Ceramic & 0.776 & 11.16 & 2.6 & 466\textsuperscript{$\dagger$} & 0.058 \\
                        \quad Connector & Various & 15.4 & - & - & 1150 & 0.088 \\
                        \quad Flex cable & Various & 33.7 & - & - & 68 & 0.026 \\
                        \quad HDI traces & Cu & 92 & 12.86 & 8.96 & 23 & 0.147 \\
                        \hline
                        \bf ROC2 LIN average  &\bf - &\bf - &\bf - &\bf - &\bf - &\bf 0.753\\
                        \quad 5 SMDs 0402 & Ceramic & 0.766 & 11.16 & 2.6 & 452\textsuperscript{$\dagger$} & 0.040 \\
                        \quad 1 SMD 0603 & Ceramic & 1.961 & 11.16 & 2.6 & 800 & 0.037 \\
                        \quad HDI traces & Cu & 65 & 12.86 & 8.96 & 23 & 0.104 \\
                    \hline
                \end{tabular}%
            \end{center}%
        \end{table}
        
        \Cref{t:estimate} gives a material budget estimate for prototype TBPX modules with a v3 HDI\@. The composition of the prototype module is specified in CMS-internal design documents, and literature radiation length values are used for the component materials. The material contents of the sensor and ROC are relatively uniform across the module, and production parameters are assumed to be well constrained. The HDI consists of a flexible PCB with three copper trace layers of 7-\SI{8}{\um} thickness, corresponding to a fractional radiation length contribution of $0.049-0.056\%$ each. The average trace density is 76.2\% across the layers, but varies strongly between the ROC0 and ROC2 linear front-end regions, with trace densities of 92\% and 64\%, respectively. The fractional radiation lengths of the ZIF connectors used for data transfer and power chaining and the data flex cable have been estimated from manufacturer documentation, with a detailed breakdown provided in \cref{s:molex_estimate}.

        Discrete components, such as connectors, surface-mount devices (SMDs) and bump bonds, were ``smeared'' across the entire surface to provide a global average estimate. The module rails cover approximately half of the pixel matrix area of the module, and extend beyond the boundaries of the pixel matrix; this area was ignored in these estimates for the purposes of this study, since only the region covered by the pixel matrix can be investigated using our method. The final average $x/X_0$ over the entire module is estimated as 0.825\%. For local regions with no contributions from SMDs, intermediate totals with and without contributions from module rails have been estimated as 0.856\% and 0.553\%, respectively. The region within the HV wirebond hole on ROC2 has an estimated $x/X_0$ of 0.394\%, with contributions only from the ROC, sensor, and bump bonds.

        Due to the significant differences in HDI components above the two regions accessible in this measurement, separate $x/X_0$ estimates have been calculated for the ROC0 and ROC2 linear regions of 0.892\% and 0.753\%, respectively.

        An estimate calculated from the material breakdown provided for the CMS Phase-0 pixel detector in Ref.~\cite{amsler2009} yields an $x/X_0$ of 0.68\% for the module stack, not including any cabling or support structures. Reference~\cite{adam2021} provides an estimate of 0.8\% for a Layer 2 Phase-1 pixel module, excluding the cable. These figures are comparable to the 0.668\% average obtained for the TBPX prototype module when neglecting the rails (support structures) and flex cable.

    \end{asection}

\end{asection}

\begin{asection}{Experimental design and preparation}%
    \label{s:design}

    Here we present a method to study the fractional radiation length topology of a tracker pixel module by the characterisation of multiple scattering behaviour in a test beam at a suitable energy range. The prototype module is positioned as the centre plane of a three-plane telescope, serving the dual purpose of device-under-test and as the central detector for three-point deflection tracking. Two MALTA (Monolithic pixel detectors from ALICE to ATLAS) prototype sensors~\cite{cardella2019,pernegger2017,snoeys2017} with $512 \times 512$ pixels of size $\SI{36.4}{\um} \times \SI{36.4}{\um}$ were used as the first and last plane due to their small thickness and material budget. The planes were separated by a nominal distance of \SI{5}{\cm}, and a scintillator directly behind the third plane provided triggering. 

    \begin{figure}[ht]
        \centering
        \subfigure[\label{f:ms_range_0825}%
                $x/X_0 = 0.825 \%$ (average over all components).%
            ]{\includegraphics[width=.47\textwidth]{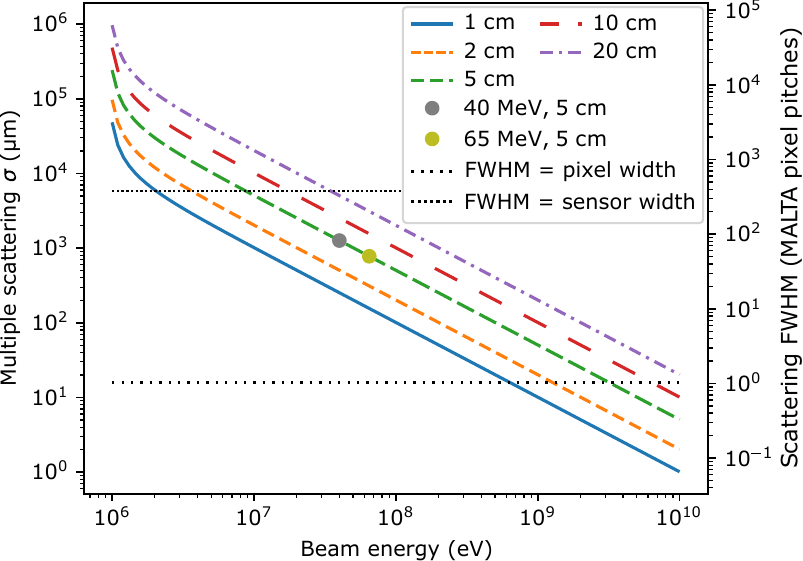}}%
        \hfill%
        \subfigure[\label{f:ms_range_0394}%
                $x/X_0 = 0.394 \%$ (ROC+sensor only).%
            ]{\includegraphics[width=.47\textwidth]{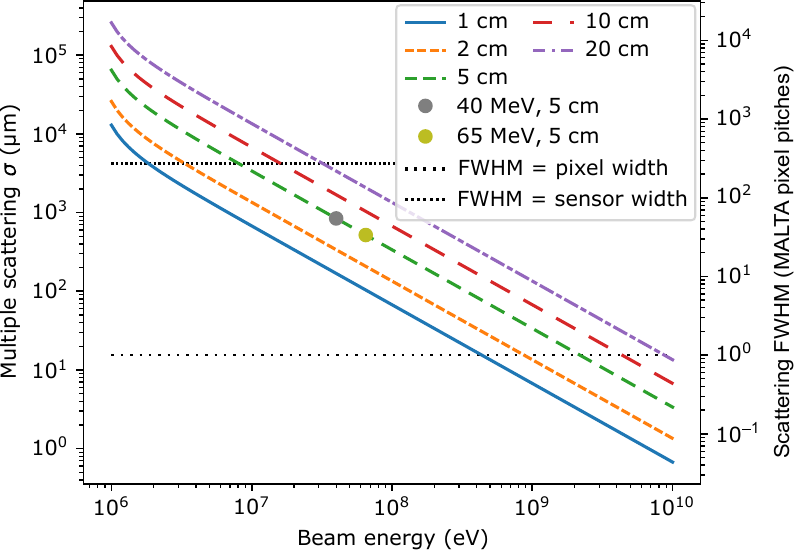}}
        \caption{\label{f:ms_range}%
            Dependence of the standard deviation of the multiple scattering distribution (in terms of transverse deflection at the third plane) on the positron beam energy, as predicted by the Highland formula, for several inter-plane spacings. Distances corresponding to a scattering distribution with full-width-half-maximum (FWHM) equal to a pixel-width and sensor-width for the final MALTA plane represent the boundaries on $\sigma$ to obtain resolvable Gaussian distributions, and are marked as horizontal dotted lines for comparison. The two energies and telescope spacings utilised in this study are highlighted.%
        }
    \end{figure}
        
    \begin{asection}{Accessible energy range for measurements}%
        \label{s:energy_range}
    
        The multiple scattering behaviour predicted by the Highland formula was investigated to determine appropriate beam and telescope parameters. A positron beam was selected due to the availability of high-purity low-energy beams of this type at the Paul Scherrer Institut (PSI) PiE1 experimental area~\cite{psi_pie1}. \Cref{f:ms_range} shows the projected standard deviation of the multiple scattering distribution on the final MALTA plane for a range of positron beam energies, for the entire module and for only the ROC and sensor (as would be observed within the HV hole on ROC2), given the estimates in \cref{t:estimate}.
        Beam energies of \SI{40}{\MeV} and \SI{65}{\MeV} were selected for study, representing relatively central points within the phase space of energies suitable for the measurement with the desired telescope spacing, as demonstrated in \cref{f:ms_range}.
    \end{asection}

        \begin{figure}[t]
            \centering\vspace{-0.5em}\hfill
            \subfigure[\label{f:threshold}%
                    Threshold distribution (target $3000 \, e^-$).%
                ]{\includegraphics[width=.42\textwidth]{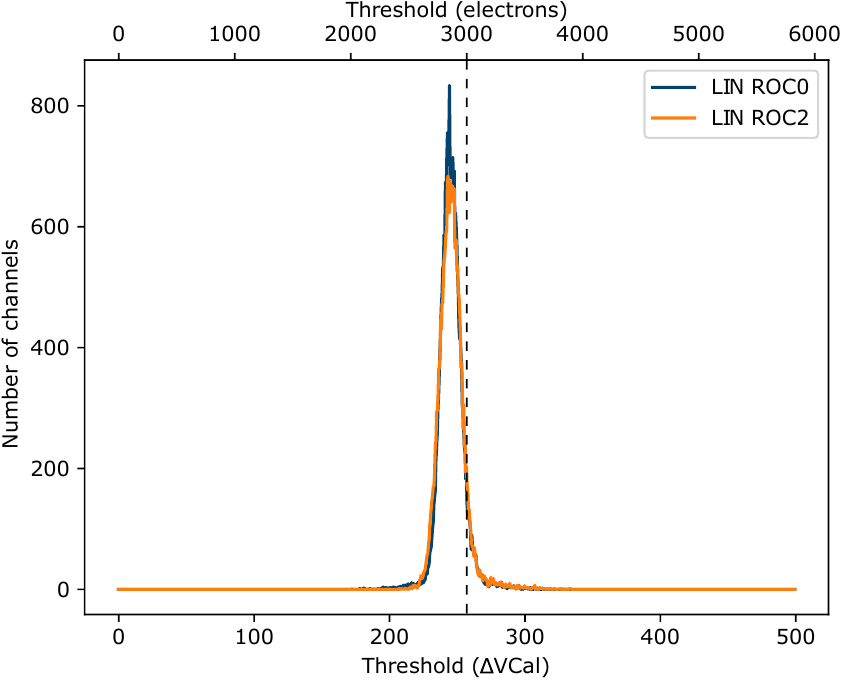}}\hfill%
            \subfigure[\label{f:noise}%
                    Noise distribution.%
                ]{\includegraphics[width=.42\textwidth]{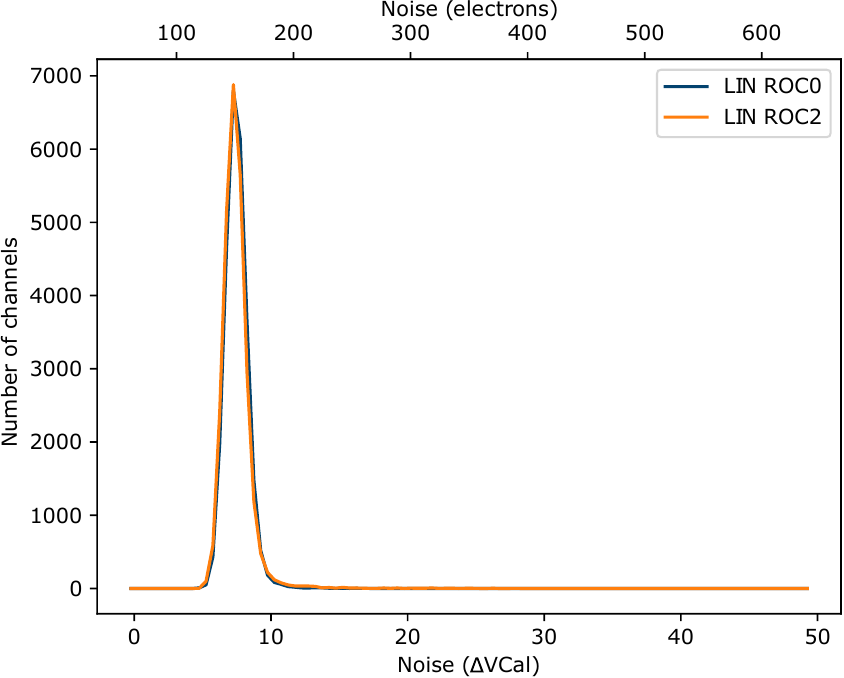}}%
            \hfill\caption{\label{f:thr_noise}%
                Threshold and noise profiles for the linear (LIN) front-end of ROC0 and ROC2. The dashed line shows the target threshold of 3000\,$e^-$.%
            }
        \end{figure}
        
    \begin{asection}{Module verification and calibration}
         
        Verification and calibration of the linear front-end regions of the prototype module prior to installation in the telescope followed the precedent of Ref.~\cite{perovic2020}. Characterisations found no noisy or dead pixels, and the threshold was tuned to a final target of $3000\,e^-$, corresponding to a calibration voltage gap of around 260\,$\Delta \mathrm{VCal}$ generating a deposited charge using the internal calibration circuitry. The conversion between $\Delta \mathrm{VCal}$ and electrons follows a linear model derived from the injection capacitance and digital-to-analogue converter resolution~\cite{dinardo2021}.
        The final threshold and noise distributions are shown in \cref{f:thr_noise}. The achieved threshold averaged slightly below the target value, but was deemed suitable for the purposes of the measurements made during this study. 

        \begin{figure}[b]%
            \centering \vspace{-0.5em}%
            \includegraphics[width=\textwidth]{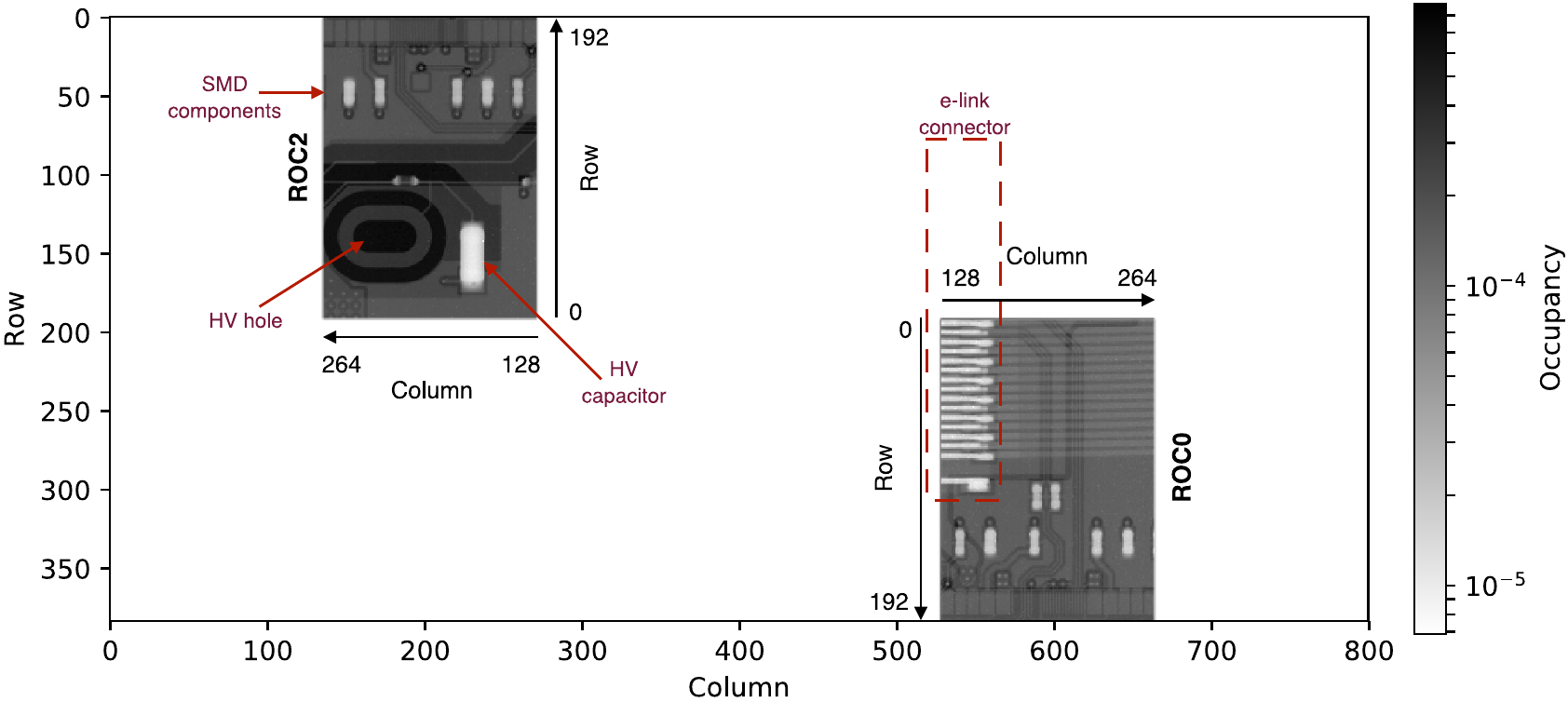}%
            \caption{\label{f:xray}%
                X-ray characterisation for the linear front-end of ROC0 and ROC2 of the prototype module performed in a Seifert X-ray chamber operating at a voltage of \SI{60}{\kV} and current of \SI{30}{\mA}. Pixel occupancies for each region are shown relative to the module frame of reference, and each ROC is annotated with the respective pixel coordinate system of the chip.%
            }%
        \end{figure}

        An X-ray characterisation was performed in a uniform X-ray flux from a Seifert X-ray chamber operating at \SI{60}{\kV} and \SI{30}{\mA}. During the test $10^8$ triggers were sent, corresponding to a mean occupancy of $\sim$\relax$10^4$ hits per pixel. The resulting map in \cref{f:xray} shows no unresponsive pixels, and allows identification of the key areas of interest on the two ROCs. The correspondence between the local coordinate system for each ROC (which will be used henceforth since each ROC is measured independently in this study) and the module layout is shown.

    \end{asection}
    
    \begin{asection}{Data acquisition and trigger logic}

        During the experiment the prototype module was read out using a modified version of the Phase-2 Acquisition and Control Framework (Ph2ACF), the standard DAQ software used to read out Phase-2 CMS tracker hardware~\cite{dinardo2021}. Data acquisition utilised a CERN FC7 FPGA board, with mounting mechanics and electrical connections as shown in \cref{f:mechanics}. The MALTA planes were each connected to a Xilinx VC707 FPGA evaluation board, and read out using the MALTA DAQ software and firmware~\cite{cardella2019}. 
        
    \begin{figure}[t]
        \centering
        \includegraphics[width=.7\textwidth]{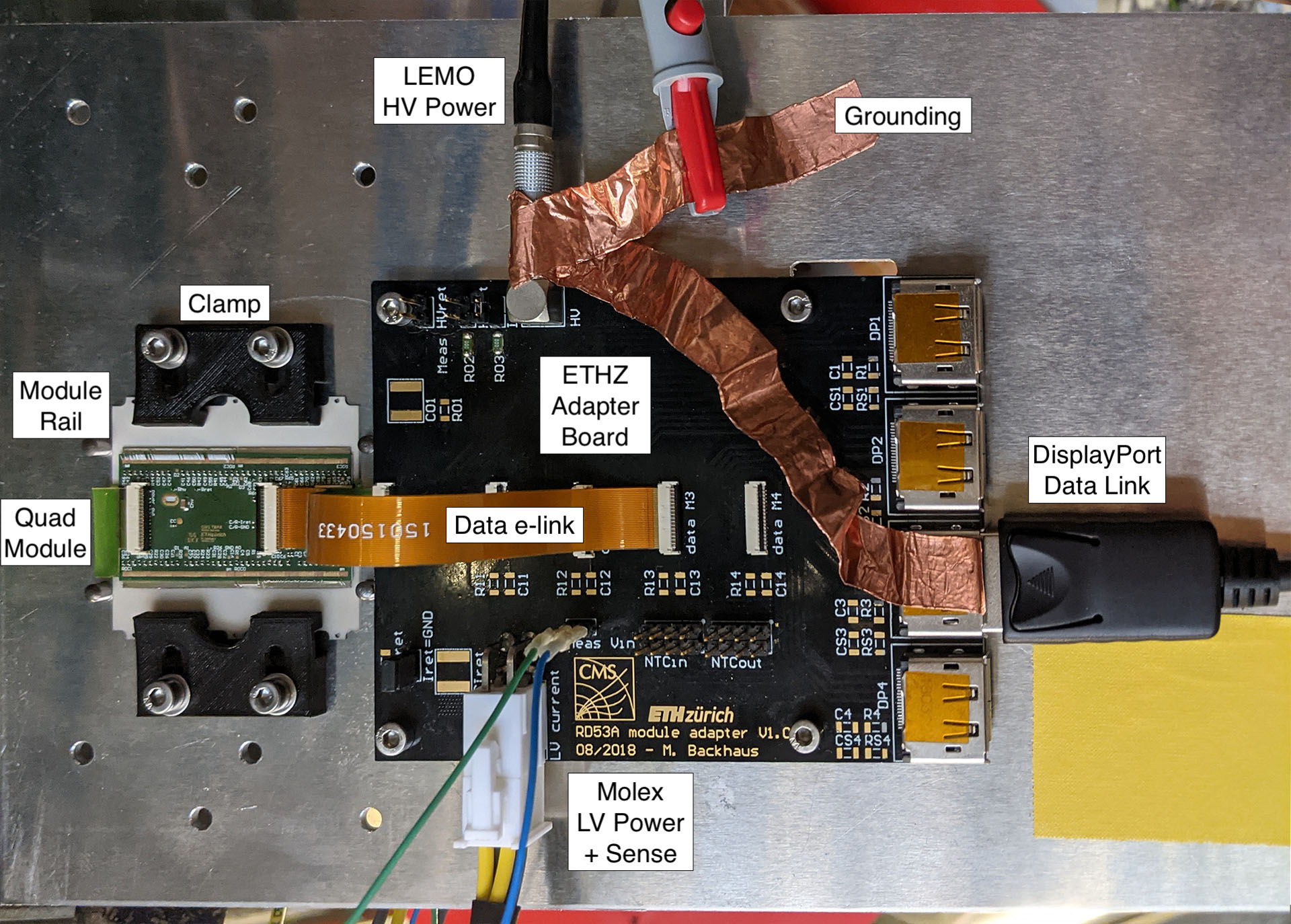}
        \caption{\label{f:mechanics}%
            Mounting mechanics for the prototype module. One ROC was exposed and measured at a time, and two sets of mounting holes allowed for the centering of each ROC within the telescope.%
        }
    \end{figure}

        Trigger signals were generated from a scintillating detector placed downstream of the telescope, and distributed to all planes using a modified implementation of the MALTA Trigger Logic Unit (TLU) on a Xilinx KC705 FPGA evaluation board. Due to the lack of a more comprehensive trigger system compatible with both the MALTA and RD53A readout systems at the time, no global trigger counter could be distributed, and trigger desynchronisation between planes posed a potential danger.
        The trigger latency for the prototype module was measured to be 825--\SI{900}{\ns}, as determined on-site using a strontium-90 source. The latency was set to 36 LHC bunch crossings, with four consecutive \SI{25}{\ns} bunch crossings read out per trigger. 
        
    \end{asection}
    
\end{asection}

\begin{asection}{Data collection and quality}%
    \label{s:data_collection}
    
    Data were collected at the PSI PiE1 facility in December 2021 at the selected beam energies of \SI{40}{\MeV} and \SI{65}{\MeV}, with a configured momentum resolution of 0.4\% FWHM (full-width-half-maximum) and a beam spot diameter of approximately \SI{5}{\mm}. Four geometric configurations (two per ROC) allowed beam coverage over a larger region of each ROC\@.
    A total of 290 million triggers was collected for the ROC2 configurations, and 340 million for ROC0.

    \begin{asection}{Event preselection}

        Events were required to contain exactly one cluster\footnote{A cluster is here defined as a maximal set of pixels above threshold in a contiguous region, where contiguous pixels are horizontally, vertically, or diagonally adjacent to another pixel in the region. These hits are assumed to originate from the passage of a single particle through the sensor.} per plane to ensure only a single valid track is present per selected event. $\mathcal{O}(10)$ noisy pixels in each MALTA plane were identified and masked prior to data collection. The MALTA sensors were configured with an active region defined as rows 120--400 and columns 140--420 in the first plane, and adjusted to match the shadow of the scintillator in the third plane (which was slightly smaller than the active region of the plane), ensuring a sharp boundary on the geometric acceptance as required for the corrections described in \cref{s:active_geometry_corrections}. The applied constraints resulted in a preselection efficiency of 7--10\% for each run, depending on geometry and beam spot quality.

    \end{asection}
        
    \begin{figure}[t]
        \centering \vspace{-1em}
        \subfigure[Data collected prior to desynchronisation produces the visible linear correlation; data after desynchronisation produces the uncorrelated background. The linear fit used to compute the coefficient of determination of the future window is shown as a dashed line.]{\label{f:correlations}%
            \includegraphics[width=.45\textwidth]{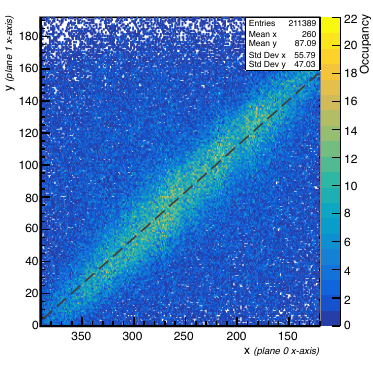}}\hspace{0.08\textwidth}%
        \subfigure[{%
        The correlation coefficient up to event $n$: ${(R^2)}^{(n)}$, and coefficient of determination of the future window $\left[ n+1, n+k \right]$: ${(R^2_{\mathrm{window}})}^{(n)}$. On desynchronisation, ${(R^2)}^{(n)}$ decays exponentially toward zero, and ${(R^2_{\mathrm{window}})}^{(n)}$ transitions rapidly to negative values.%
            }]{\label{f:r2}%
            \includegraphics[width=.45\textwidth]{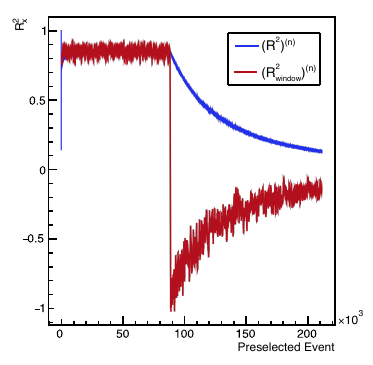}}%
        \caption{\label{f:synchronisation}%
            Correlation plot and coefficients for correlations between the $x$-axes of plane 0 and plane 1 in run 911 (ROC2,  \SI{40}{\MeV} beam energy), which suffered desynchronisation after $\sim88\cdot10^3$ preselected events.%
        }
    \end{figure}

    \begin{asection}{Trigger synchronisation}%
        \label{s:sync}

        The lack of a global trigger counter or tag communicated by the TLU to all planes made it possible for the event data streams to desynchronise upon loss of a single trigger (or sequence of triggers) in one of the three planes. This results in data corruption due to events consisting of mismatched triggers, and was observed in $\sim$29\% of runs taken, corresponding to over 182 million events (prior to the above mentioned pre-processing constraints). An offline algorithm utilising correlations between the spatial positions of clusters in the planes was developed to detect and correct desynchronisation. For each set of correlated axes between a pair of planes, the correlation coefficient ${(R^2)}^{(n)}$ was computed for each event $n$ cumulatively as
        \begin{equation}
            {(R^2)}^{(n)} = \frac{{\left(S^{(n)}_{xy}\right)}^2}{S^{(n)}_{xx} \cdot S^{(n)}_{yy}},
        \end{equation}
        where $S^{(n)}_{ab} = \sum_{i = 1}^n (a_i - \bar a) (b_i- \bar b)$ is the sum of the products of the differences from the mean of all hits up to $n$ in axes $a$ and $b$, and $x$ and $y$ represent one pair of correlated axes.
        A linear fit to the current correlation was then compared to a window of $k$ future events to determine the coefficient of determination of the window:
        \begin{equation}
            {(R^2_\text{window})}^{(n)} = 1 - \sum_{i=n+1}^{n+k} \frac{{(y^{(i)} - \hat y^{(i)})}^2}{{(y^{(i)} - \bar y_\mathrm{window}^{(n)})}^2}.
        \end{equation}
        Here $y^{(i)}$ gives the position on the second correlated axis $y$ for an event~$i$ within the window, $\hat y^{(i)}$ the expected position on axis $y$ given the linear fit and position on the first correlated axis $x$, and $\bar y^{(n)}_\mathrm{window}$ the mean position on axis $y$ for events within the window. The correlations and coefficients for an example run are shown in \cref{f:synchronisation}.

        A normalised indicator henceforth referred to as the \emph{stability} was defined as 
        \begin{equation}
            s^{(n)} = \frac{1}{{(R^2)}^{(n)}} \left< \frac{d{(R^2)}^{(n)}}{dn} \right>_\mathrm{window} \approx \frac{{(R^2_\mathrm{window})}^{(n)}-{(R^2)}^{(n)}}{{(R^2)}^{(n)}},
        \end{equation}
        and proved a robust indicator of desynchronisation, with local variations generally bounded by $\pm0.5$ about a nominal stable equilibrium at $\bar s = 0$ for well-correlated data. On desynchronisation, a rapid transition to a new equilibrium centred at approximately $\bar s^* = -2$ is observed, and the variance of the stability increases as more uncorrelated data is added, as may be seen in \cref{f:stability} for an example run. On deterioration of $s$ at an event~$n$ such that $\bar s^{(n)} - s^{(n)} > t$ for an adjustable threshold $t$, desynchronisation is assumed to have occurred later than event $n-k$, since the transition time for $s$ between the two equilibria is at most the window width $k$. Within this study, $t$ was chosen as $1$, and a window width of $k=500$ was used.

        \begin{figure}[t]
            \centering \vspace{-1em}
            \subfigure[Stability evolution prior to resynchronisation.]{\label{f:stability_pre}%
                \includegraphics[width=.49\textwidth]{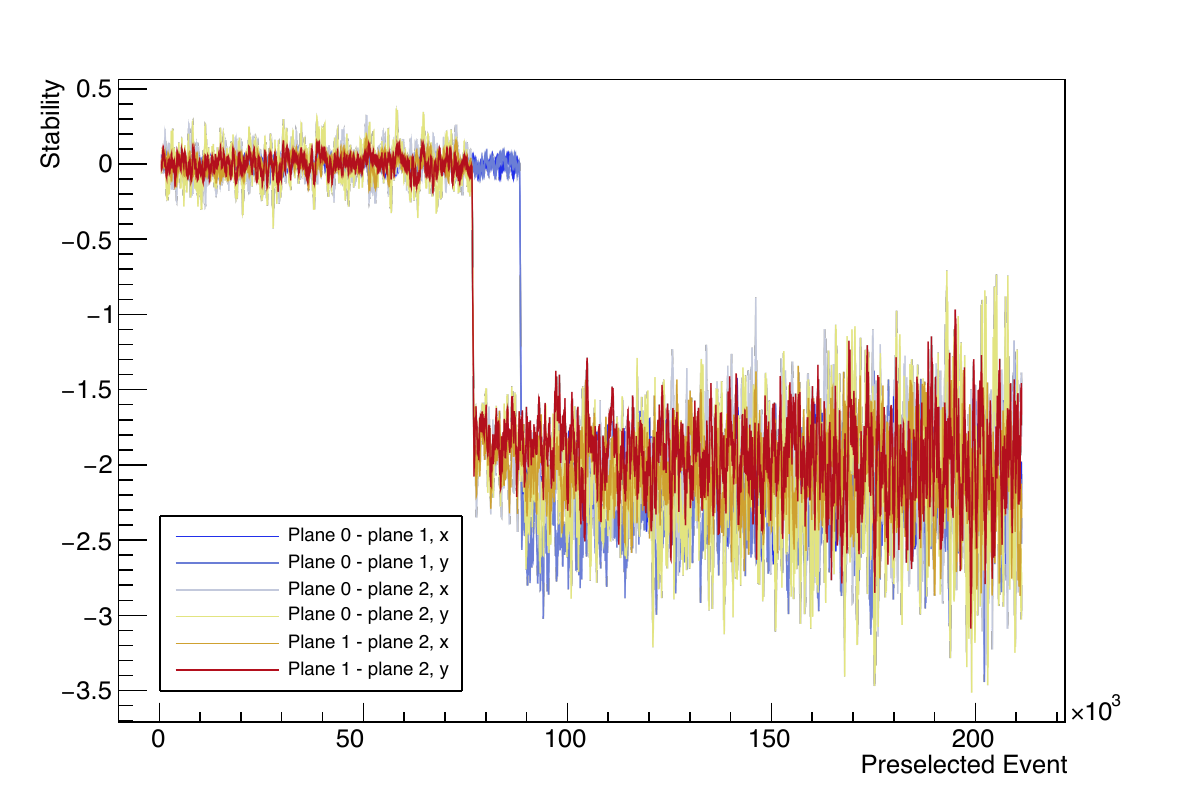}}\hfill%
            \subfigure[Stability evolution after resynchronisation.]{\label{f:stability_post}%
                \includegraphics[width=.49\textwidth]{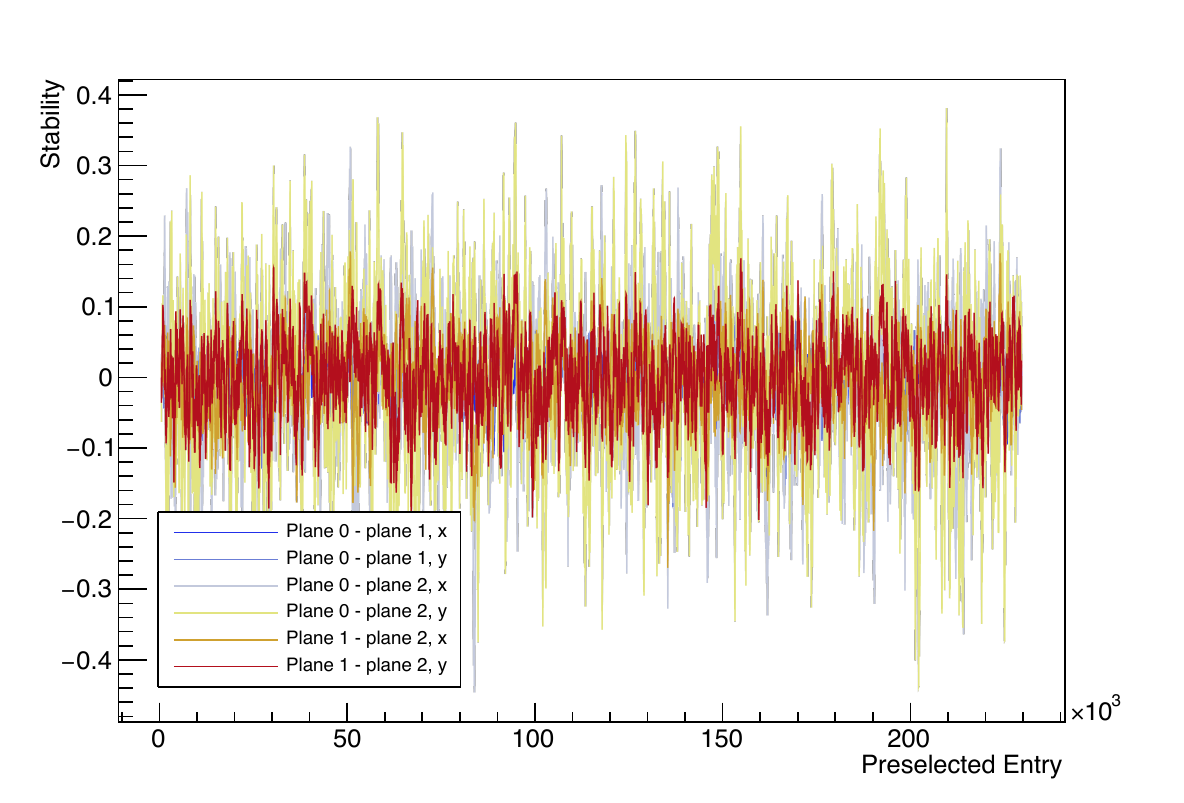}}%
            \caption{\label{f:stability}%
                Evolution of the stability $s$ for all six correlations of run 911 (ROC2,  \SI{40}{\MeV} beam energy). Desynchronisation of all four correlations involving plane 2 occurs near entry $75\cdot10^3$, and of the remaining correlations between plane 0 and plane 1 near entry $88\cdot10^3$. These desynchronisations are successfully recovered by the resynchronisation process.%
            }
        \end{figure}

        An algorithm was developed to recover synchronisation by offsetting the trigger number in the plane suffering desynchronisation forward or backward. First, correlations are scanned until desynchronisation is detected at an event $n$. All events up to event $n-k$ are marked as synchronised. Synchronisation recovery is attempted on events starting from event $n+k$. The following is repeated for $r = \pm i \cdot c$ where $i \in \mathbb Z$ up to some maximal number of retries $i_\text{max}$ and $c$ is an offset multiplier: 
        \begin{enumerate}
            \item Reindex triggers in the desynchronised plane as $n \rightarrow n + r$.
            \item Compute the next window starting from event $n+k$, and check if $s^{(n+k)}$ meets the synchronisation criterion. If so, keep the current reindexing, and continue the scan. Otherwise retry with the next value of $r$.
        \end{enumerate}
        
        The algorithm achieved an 18.8\% increase in total yield, with improvements variable between energies and geometries, dependent on the extent and type of desynchronisation. Recovery of synchronisation succeeded for 96\% of all desynchronisations involving plane 0 or 2, with the remainder occurring too early. No desynchronisations of plane 1 could be recovered. No biases or inconsistencies were observed in the recovered datasets.

        The main cause of desynchronisation in planes 0 and 2 was an issue with the MALTA DAQ software at the time, in which corrupted data packets with the bunch crossing counter equal to 0 could trigger an overflow of the 12-bit counter. Since the MALTA DAQ system only transmitted data for triggers with hit pixels (and thus empty events were inserted for missing bunch crossing counter values), this resulted in the insertion of 4096 empty events for the affected plane. In these cases, the algorithm was successfully used with values of $c = 4096$ and a maximum of 128 retries to resynchronise most runs affected by this.

        In a small number of cases, plane 1 desynchronised from the others, and all cases of this occurred very early in the run, within the first O\relax(10\,--\,100) events. This prevented the determination of a baseline stability value for the algorithm and led to the resynchronisation process failing on these datasets.
    \end{asection}

\end{asection}

\begin{asection}{Analysis}%
    \label{s:analysis}
        
    The alignment of the planes was extracted by minimisation of residuals within the combined dataset for each geometry. Alignment for a small sample of runs was cross-checked with the \emph{coarse alignment} method in the Proteus analysis framework~\cite{proteus}. No significant deviations between alignment methods were noted, and rotations of planes about the beam axis were $\ll \ang{1}$ in all cases.

    The analysis of the multiple scattering data centred on two types of deflection angle extracted from the datasets: the phi-invariant deflection angle in the global coordinate system, where phi is defined as the angle about the beam axis, and the projection of the deflection on the $x$-$z$ and $y$-$z$ planes, where the $x$ and $y$ axes are taken from the local coordinate system of the central plane and the positive $z$-axis is defined in the direction of the beam.

    To provide a position-resolved map of $x/X_0$, the region of interest of the ROC forming the central plane was divided into rectangular or polygonal subregions, and the extracted angles were separated based on the subregion the particle had scattered in. For each subregion and extracted angle type an appropriate distribution (as described in \cref{s:fit_strategy}) was fitted to a histogram of matching deflections. The results of these separate fits were later combined to a single value as described in \cref{s:fit_combination}.

    \begin{asection}{Extraction of angles and fit strategy}%
        \label{s:fit_strategy}
        
        Fits were performed using the RooFit framework~\cite{roofit}. A variable binning of 40--80 bins per sub-region ensuring a minimal average occupancy of 250 entries per bin was adopted to ensure sufficient statistics, and no fit was attempted in ROC subregions with less than $10^4$ events.

        The projected angle was computed directly from the two track segments (between the first and second plane, and second and third plane). The resultant distribution matched the expected features of the Moli\'ere distribution, with a Gaussian core and two single-scatter tails $\propto 1/\sin^4\theta$. A fit to a double-sided crystal ball function (DSCB) consisting of a Gaussian core with two power-law tails with variable exponent $N \approx 4$ was utilised~\cite{crystalball}. 

        The global angle was computed from the track segments using the cosine formula, and was introduced in an attempt to mitigate the susceptibility of the projected angle fitting to some of the issues described in the following sections. The expected distribution for the global angle does not follow that of the projected angles, and is instead derived by assuming $\theta_x$ and $\theta_y$ to be approximately independent Gaussian random variables $\mathcal N(\mu, \sigma)$ with mean $\mu$ and width $\sigma$ for a range of the global angle constrained to small values, giving $\theta^2 \approx \theta_x^2 + \theta_y^2$. Additionally it is assumed that the mean projected scattering angle $\mu$ is $0$, allowing the sums of squares of the $N$ identical $\mathcal N(0, \sigma)$ distributions to be described as a single sigma-weighted chi-squared distribution $\sigma^N \cdot \chi^2(k=N)$ with $N$ degrees of freedom. A chi-squared distribution is a special case of the more general gamma distribution $\Gamma(\gamma, \beta)$ (with shape parameter $\gamma$ and scale parameter $\beta$), with defining relation $\chi^2(k) = \Gamma(\frac{k}{2}, 2)$. The distribution of the square of the global scattering angle may thus be described as follows:

        \setlength\abovedisplayskip{-12pt}%
        \begin{align}%
            \label{e:norm}\theta_k &\sim \sigma \cdot \mathcal{N}(0, 1), \\
            \label{e:chi2}\theta_k^2 &\sim \sigma^2 \cdot \chi^2(1) = \sigma^2 \cdot \Gamma\left(\frac12, 2\right) = \Gamma\left(\frac12, 2 \sigma^2\right), \\
            \theta^2 &\approx \theta_x^2 + \theta_y^2 \sim \Gamma(\gamma = 1, \beta = 2 \sigma^2),
        \end{align}%

        \noindent where in \cref{e:norm,e:chi2} the angular distribution $\theta_k$ generalises to both $\theta_x$ and $\theta_y$.

        However, $\theta_x$ and $\theta_y$ are not truly independent as they are related via a polar symmetry within the underlying scattering mechanism. The condition on $\gamma$ may be relaxed to allow the fitting method to select a suitable number of effective degrees of freedom. To ensure the fitted value of $\sigma$ scales correctly with $\gamma$, the expectation value of the distribution must be maintained. Using the expectation value of the gamma distribution as given in \cref{e:gamma_prop}, the relation between $\sigma$ and $\beta$ is refined as follows:

        \setlength\abovedisplayskip{-12pt}
        \begin{gather}%
            \label{e:gamma_prop}%
            \mathrm{E}\left[\Gamma(\gamma, \beta)\right] = \gamma \cdot \beta, \\
            \mathrm{E}\left[\Gamma(\gamma, 2\sigma^2/\gamma)\right] = 2\sigma^2.
        \end{gather}%

        \noindent Thus both $\gamma \in [0.5, 1]$ and $\beta$ can be fitted, and $\sigma$ is extracted by the relation $\sigma = \sqrt{\frac12 \gamma\beta}$. To ensure the validity of the small angle approximation, $\theta^2$ is constrained at an upper limit of $\theta^2 \le {(\sigma_x + \sigma_y)}^2$, where $\sigma_x$ and $\sigma_y$ are the standard deviations of the projected distributions. 

    \end{asection}

    \begin{asection}{Active geometry corrections}%
        \label{s:active_geometry_corrections}
        
        \begin{figure}[t]
            \centering
            \includegraphics[width=.72\textwidth]{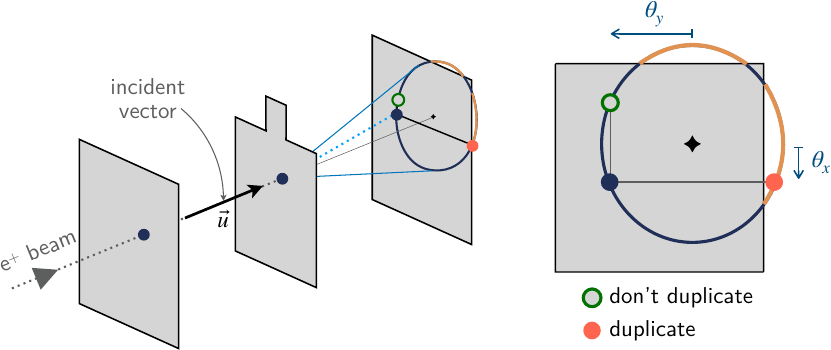}
            \caption{\label{f:proj_ag_corr}%
                To build the active region correction for projected angles, a cone of all possible deflections corresponding to the same incident vector and global deflection angle is drawn. For each axis, projected angles are duplicated with opposite sign if the new projected angle would fall outside of the active region of the final telescope plane (orange). Otherwise, they are not (blue).%
            }
        \end{figure}

        Corrections to both the projected and global angle distributions were required in order to account for inefficiencies in particular angle bins due to the spatial limits of the final telescope plane. In order to do so, a geometric fiducial region was defined by considering the \emph{active region} of the final plane. This was generally defined by a combination of the limits of the active area within the MALTA plane pixel matrix, and the shadow of the scintillator used for triggering on the final plane. Limits were chosen conservatively to ensure clean boundaries of the active region. 

        In the case of the projected angle distribution, a discrete approach was taken on a per-event basis. As shown in \cref{f:proj_ag_corr}, a cone of all possible deflections sharing the same incident vector $\vec u$ and global deflection angle $\theta$ is drawn. For each projection plane, a check is made to determine if the other projected angle within the same plane intersecting the cone is within the active region. If it is, no correction is made. If not, this second angle is counted to account for the possibility that such a deflection occurred but was not detected since it was outside of the region of acceptance of the final plane. Since each polar deflection angle on the cone is equally likely, this method ensures fair counting of deflections for which only one of the two possibilities would be detected, and thus prevents undercounting of large deflection angles for hits close to the edges of the centre plane.

        \begin{figure}[b]
            \centering
            \includegraphics[width=.72\textwidth]{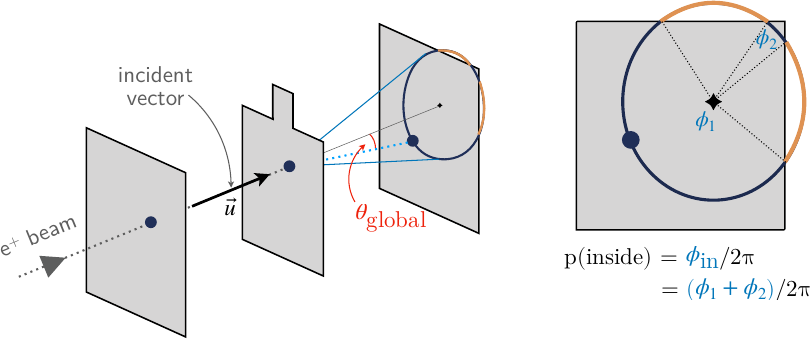}
            \caption{\label{f:glob_ag_corr}%
                In the case of the active region correction for global angles, the angles in the plane perpendicular to the cone axis subtended by arcs describing trajectories within the active region of the final plane are summed to give the probability of any event with the same incident vector and global deflection angle being accepted by the telescope. In the case pictured, the probability is given as $p_\text{inside} = (\phi_1 + \phi_2)/2\pi$.%
            }
        \end{figure}

        The global angle correction was implemented as a weighting of each event by the inverse of the probability of a deflection with the same incident vector and global angle traversing the active region of the final plane. As demonstrated visually in \cref{f:glob_ag_corr}, a cone with opening angle $\theta$ and axis corresponding to the incident vector is intersected with the active region of the final plane. The sum of subtended polar angles for arcs inside the active region within the plane of reference perpendicular to the cone axis is then divided by $2\pi$ to give the probability that any deflection with the same incident vector and global angle would be within the acceptance of the telescope.

    \end{asection}
    
    \begin{asection}{Edge region corrections}%
        \label{s:edge_region_correction}

        A further set of corrections was necessitated by the conical beam spread through the telescope, as shown in \cref{f:er_corr}. Since the beam is not parallel, scatters through subregions close to the edge of the central plane are subject to a shadow in the centre of the distribution.  The non-parallel nature of the beam causes undercounting of small-angle scatters close to the edge of the centre plane since they will fall outside the active region of the final plane, compared to large angle scatters, which are correctly accounted for by the 2D angle correction described in \cref{s:active_geometry_corrections}.

        \begin{figure}[b]
            \centering
            \includegraphics[width=\textwidth]{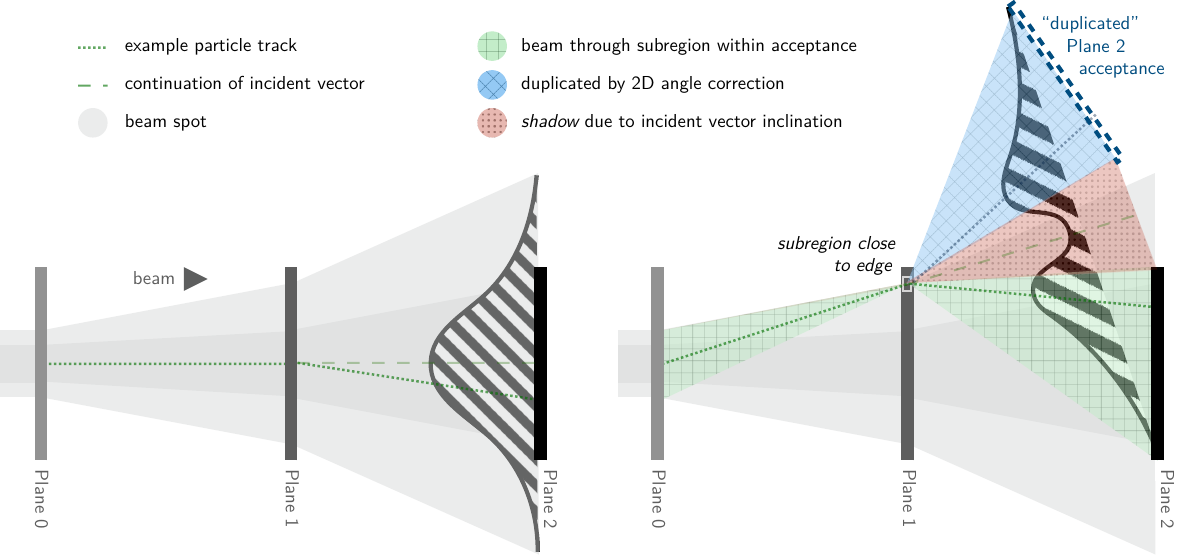}
            \caption{\label{f:er_corr}%
                Effect suppressing small angle scatters in edge regions of the centre plane, and the impact on the projected angle distributions observed. The original data distribution is represented by the green shaded area, and is effectively duplicated to the blue area by the active region correction as described in \cref{s:active_geometry_corrections}. The red region corresponds to possible trajectories excluded by the non-horizontal nature of the incident vector.%
            }
        \end{figure}
        
        \begin{figure}[tpb]
            \centering
            \includegraphics[width=\textwidth]{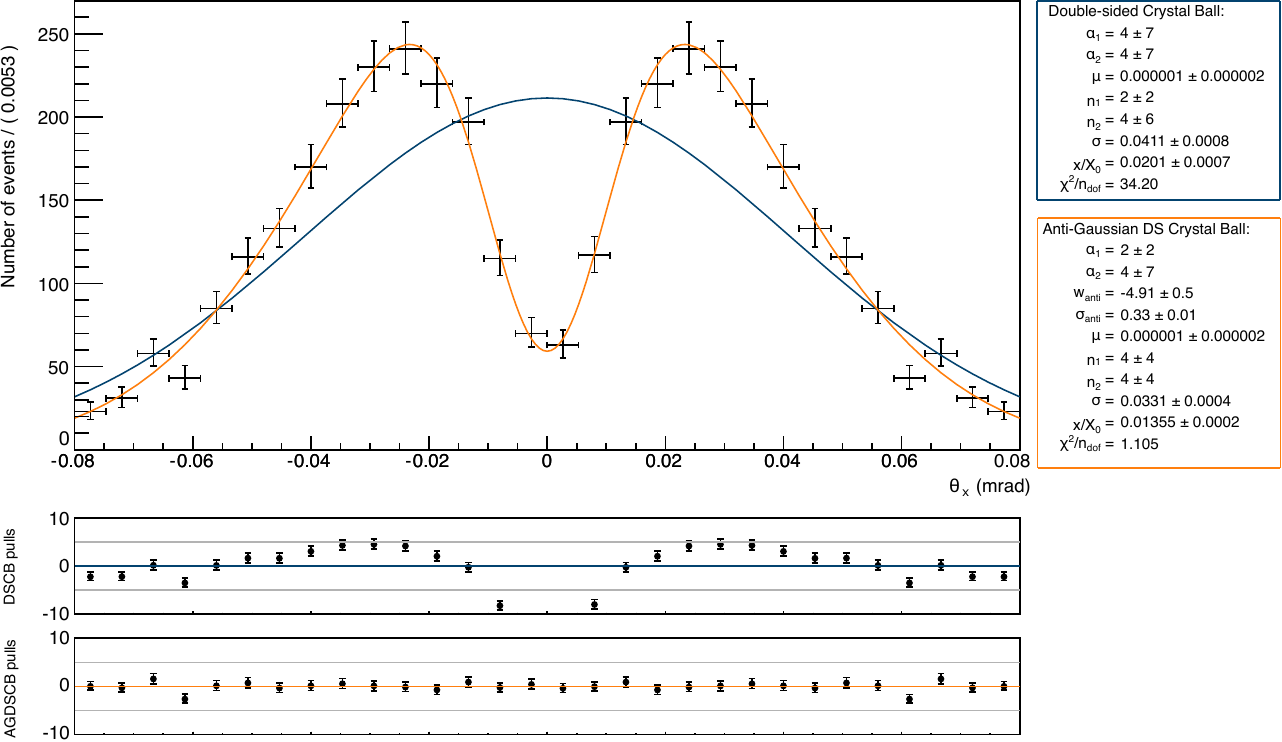}
            \caption{\label{f:er_fits}%
                $\theta_x$ distribution for a subset of ROC2, \SI{40}{\MeV} data, within the single subregion defined by column $\in [136, 144]$ and row $\in [24, 36]$. DSCB (blue) and AGDSCB (orange) fits are compared, and fitted parameters and pulls are shown, as well as the statistical uncertainty in each bin. The overestimated $x/X_0$ value of 2.01\% for the DSCB fit corrects to 1.36\% in the AGDSCB fit.%
            }
        \end{figure}

        \begin{figure}[tpb]
            \centering
            \subfigure[AGDSCB fit on $\theta_x$]{\label{f:agdscb}%
                \includegraphics[height=.225\textheight]{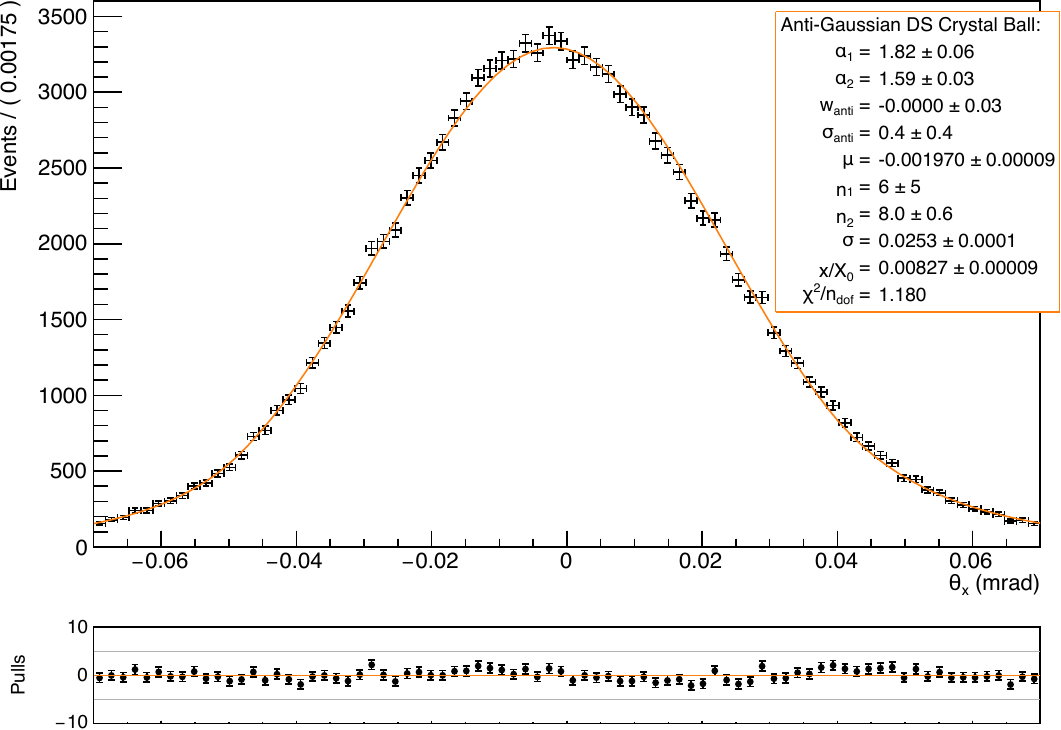}}
            \hfill
            \subfigure[GammaPlus fit on $\theta$]{\label{f:gammaplus}%
                \includegraphics[height=.225\textheight]{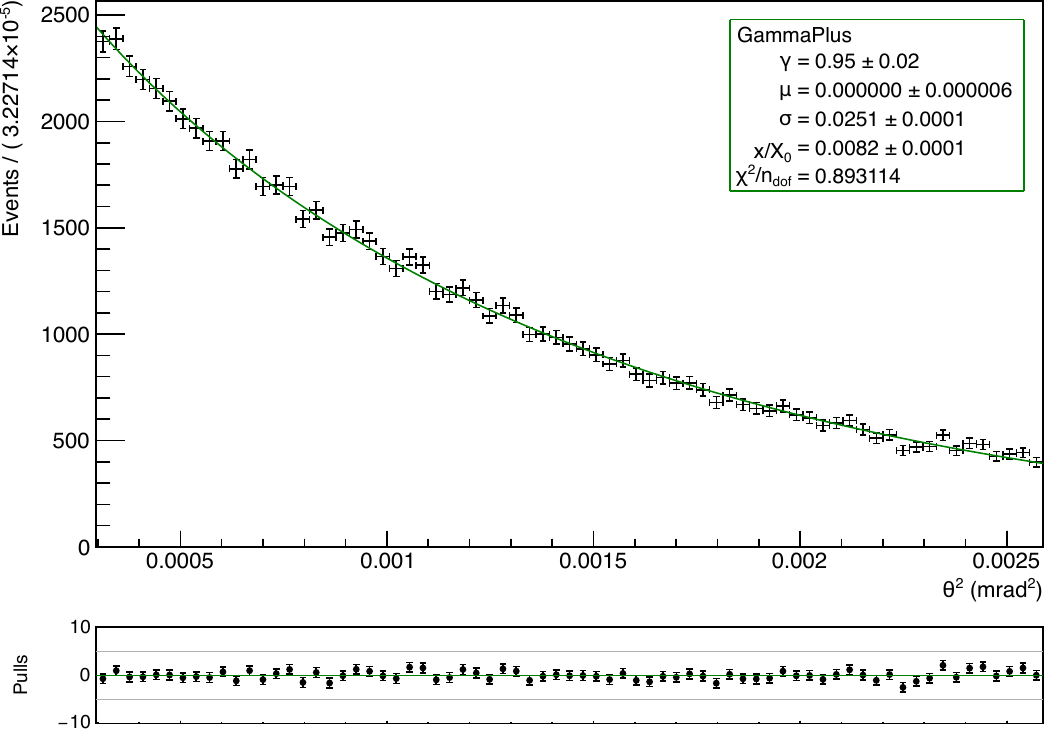}}
            \caption{\label{f:fits}%
                $\theta_x$ and $\theta$ distributions, and corresponding AGDSCB and GammaPlus fits, for a subset of ROC2, \SI{40}{\MeV} data, within the single subregion defined by column $\in [192, 200]$ and row $\in [108, 120]$.
            }
        \end{figure}
        
        In the case of the projected angle, a double-peaked distribution forms for subregions close to the edge of the central plane, as is shown schematically in \cref{f:er_corr}. The effect of this is shown in \cref{f:er_fits}, and leads to an overestimate of the fractional radiation length due to the peak of the distribution being fit to lower values.
            
        The inefficiency observed at small angles can be reasonably modelled as a suppressive Gaussian contribution (due to the Gaussian nature of the beam). In order to correct for the effect, a composite probability distribution function (PDF) was devised consisting of the summation of a DSCB distribution with a variable weight negative Gaussian core sharing the mean of the former and restricted to $\sigma_\text{gaus} \in [0.1, 1.2] \cdot \sigma_\text{DSCB}$. The PDF was clipped to positive values only, and is henceforth referred to as an Anti-Gaussian + DSCB (AGDSCB). 

        The global angle distribution is pulled downward for small angles in these regions by the edge effects described above. As a result, the distribution generally becomes non-monotonically-decreasing (as would be required of a $\Gamma$-distribution with $\gamma \in [0.5, 1]$). The issue was partially mitigated by removing all bins up to $N_\text{bins}/10$ bins past the maximal value of the global angle distribution. A careful manual optimisation identified this choice as successfully restricting the fit to a range that is well modelled by a $\Gamma$-distribution, and yielded values of $\sigma$ in close agreement with that extracted from projected angle fits. The improved range selection and fit method is collectively termed GammaPlus. Example AGDSCB and GammaPlus fits for a region near the centre of ROC2 for data taken at \SI{40}{\MeV} are shown in \cref{f:fits}.

    \end{asection}
    
    \begin{asection}{Combining projected and global angle fits}%
        \label{s:fit_combination}
        After the fitting of $\sigma$ and the application of the inverse Highland formula for each of the $\theta_x$, $\theta_y$ and $\theta$ distributions, these were combined as though they were equal-weight independent measurements, despite the caveat of non-trivial correlations between the global and local angles. For each value, a minimal fit quality cut corresponding to $\chi^2/\text{ndf} < 2$ was applied, and only values of $x/X_0$ extracted from fits meeting these criteria were combined. 
    \end{asection}

    \begin{asection}{Uncertainty estimates}%
        \label{s:uncertainty_esimate}
        Uncertainties on the value of $\sigma$ for each distribution were taken directly from the output of \texttt{Hesse} within RooFit, which inverts the second derivative matrix for the fit. These were combined with the uncertainty from multiple scattering in air described below, and then propagated through the inverse Highland formula in combination with the beam energy uncertainty to give a symmetric uncertainty on $x/X_0$ for each subregion. 

        The expected worst-case spatial resolution of the MALTA planes, assuming all clusters only contain one hit, is $36.4/\sqrt{12} \approx 10.5\,\si{\um}$. Since the Gaussian $\sigma$ of the scattering distribution for the smallest expected value of $x/X_0$ is $\sim$\relax$\SI{519}{\um}$ as shown previously in \cref{f:ms_range_0394}, the mean expected uncertainty on each angle for this region from the MALTA plane resolution is 2\% in this worst-case region, and much less in regions of greater material content. Since this is small compared to the other uncertainties mentioned here, it has been neglected.

        \begin{description}
            \item[Beam energy spread]~\\
            The beam energy loss and spread in the first telescope plane were computed from the Landau-Vavilov distribution for each beam energy, and the estimated beam momentum spread of 0.8\% FWHM\footnote{This value was taken as a conservative estimate for a nominal bound configured to 0.4\% due to difficulties in setting such narrow acceptances correctly.}~\cite{vavilov1957,landau1944,bichsel1988}. The most probable energy loss was subtracted from the nominal beam energy, and the momentum resolution and energy spread were combined to form the uncertainty on the beam energy. The energy spread caused by the first telescope plane contributed 5\% of the total energy uncertainty for the \SI{40}{\MeV} configuration and 3\% for the \SI{65}{\MeV} configuration, indicating that the width of the beam momentum band comprised the dominant energy uncertainty contribution, rather than the material content of the first plane.
                
            \begin{figure}[t]
                \centering
                \includegraphics[width=.98\textwidth]{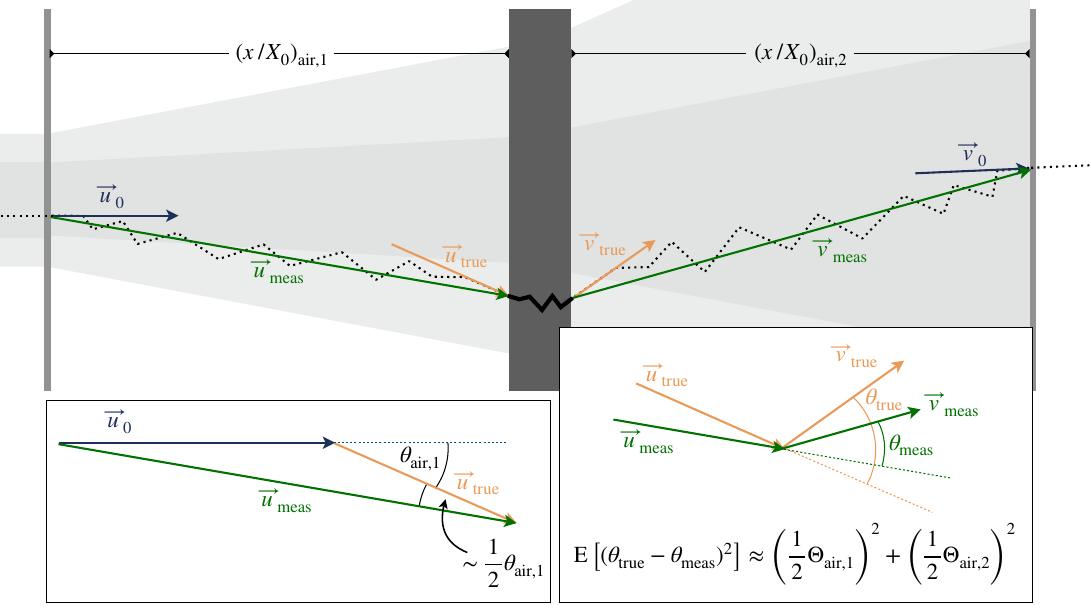}
                \caption{\label{f:rad_length_air}%
                    The effect of the multiple scattering in regions between planes. The air is separated into two regions and characterised by the fractional radiation length ${(x/X_0)}_\mathrm{air}$ of each, with a corresponding RMS scattering angle $\Theta_\mathrm{air}$. For the event shown, $\theta_\mathrm{true}$ gives the true scattering angle through the centre plane, and $\theta_\mathrm{meas}$ gives the angle measured in the analysis. The transverse extent of the scattering is visually exaggerated.%
                }
            \end{figure}
        
            \item[Multiple scattering in air]~\\%
            \label{s:ms_in_air}%
            Multiple scattering of the traversing particle in air within the telescope results in an offset from the true extracted angle, as shown by the geometric arguments in \cref{f:rad_length_air}. Assuming the distribution of the multiple scattering in each region of air to be Gaussian with RMS scattering angle $\Theta_{\mathrm{air},i}$, the offset follows a Gaussian distribution with variance

            \begin{equation}
                \label{e:air_unc}
                \mathrm{E} \left[ \Delta\theta^2 \right] \approx \sum_{i=1,2} {\left( \frac12 \Theta_{\mathrm{air},i} \right)}^2.
            \end{equation}

            The value of $\Theta_{\mathrm{air}, i}$ was computed from the radiation length of dry air at \SI{1}{\atm}, taken as \SI{36.62}{\gram/\cm^2}~\cite{pdg2022}. At \SI{40}{\MeV}, \SI{5}{\cm} of air produces a scattering angle distribution of width \SI{2.7}{\milli\radian}, corresponding to a total uncertainty from air of \SI{1.9}{\milli\radian} from \cref{e:air_unc}. Compared to an expected scattering distribution of \SI{28}{\milli\radian} for an area of an $x/X_0$ of 1\%, this constitutes a $7\%$ uncertainty on the scattering angle. After propagation through the Highland formula, the multiple scattering in air comprised the dominant uncertainty contribution in all ROC subregions, constituting over 75\% of the total uncertainty in subregions of moderate material content $x/X_0 \lesssim 1\%$.
        \end{description}

    \end{asection}

\end{asection}

\begin{asection}{Results}%
    \label{s:results}

    Two different ROC subregion structures were used to group hits within the centre plane before fitting regular rectangular subregions of dimension $8 \times 12$ pixels, and polygonal subregions defined as regions of relatively uniform material composition (based on the X-ray measurements in \cref{f:xray}).

    \begin{asection}{Rectangular subregion fractional radiation length maps}%
        \label{s:rectangle_results}
        
        \begin{sidewaysfigure}[p]
            \centering 
            \includegraphics[width=.94\textwidth,page=1]{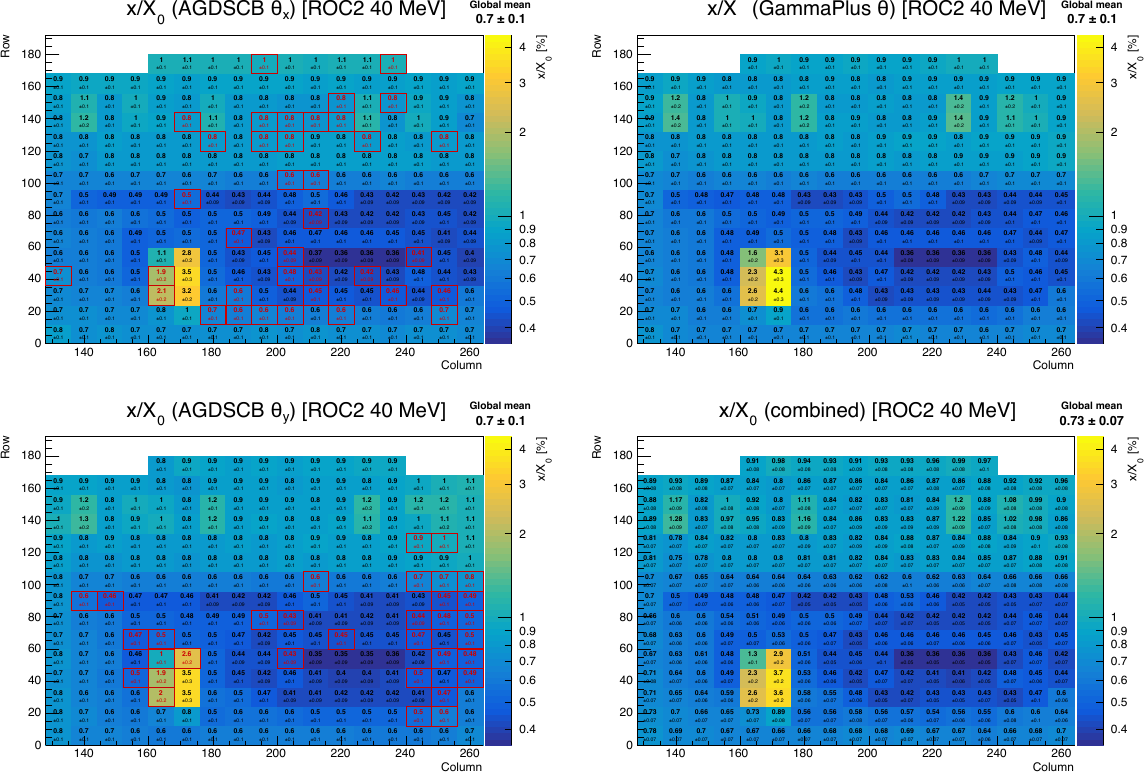}
            \vspace{-0.5em}
            \caption{\label{f:roc2_40}%
                Fractional radiation length map of ROC2 at \SI{40}{\MeV} for all three fitting methods, and the combined value. Empty subregions were not fitted due to insufficient statistics, and subregions outlined in red did not meet the quality criteria for inclusion in the combined measurement. The colour axis is identical for all methods.%
            }
        \end{sidewaysfigure}
    
        \begin{figure}[b]
            \centering
            \includegraphics[width=.88\textwidth,page=2]{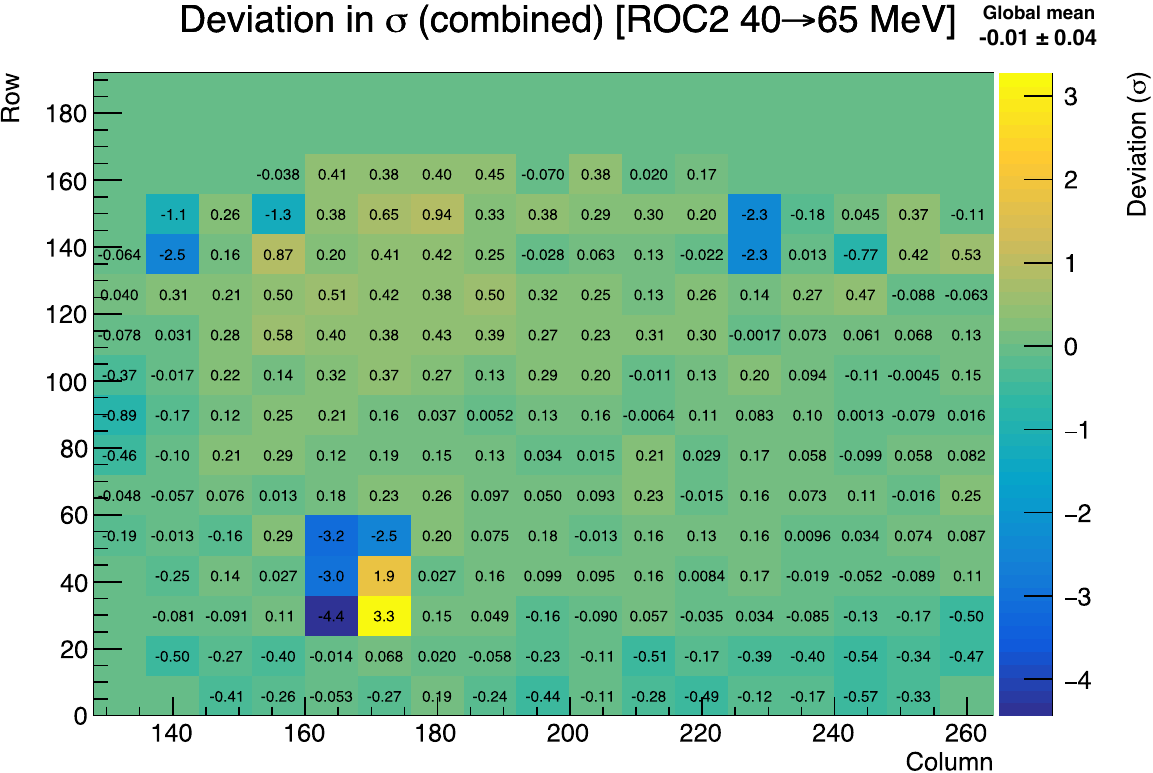}
            \caption{\label{f:compare_roc2}%
                Deviation (given in number of standard errors) between 40 and \SI{65}{\MeV} combined results for ROC2. Subregions without numeric value were not fitted in one or more of the datasets due to insufficient statistics.%
            }
        \end{figure}
    
        \begin{figure}[t]
            \centering
            \includegraphics[width=.88\textwidth,page=1]{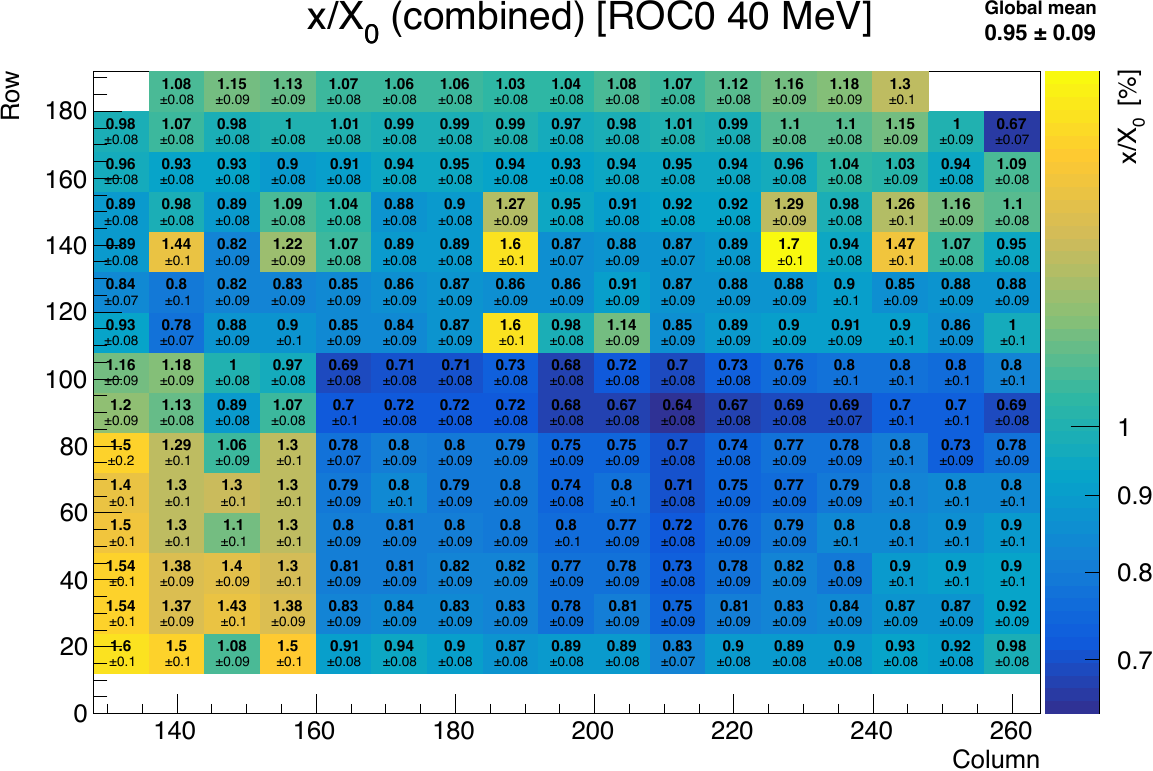}
            \caption{\label{f:roc0_40_c}%
                Combined value radiation length map of ROC0 at \SI{40}{\MeV}. Empty subregions were not fitted due to insufficient statistics. Inflation of $x/X_0$ by up to 0.2 percentage points due to edge effects accentuated by the use of a narrower beam compared to the ROC2 measurements is visible in the outermost subregions.
            }
        \end{figure}
        
        \Cref{f:roc2_40} shows maps for the ROC2 \SI{40}{\MeV} dataset for all fitting methods, and the combined value. The fitting methods generally agree very well, with only regions above the SMD components and a small number of subregions on the right side of the AGDSCB fit for $\theta_y$ deviating by more than one standard error between the methods.

        The \SI{40}{\MeV} data are in agreement with the \SI{65}{\MeV} dataset within one standard deviation in all regions except above the SMD components, and a comparison is provided in \cref{f:compare_roc2}. 
        The stronger deviations above SMD components are attributed to the sharp edge between regions of vastly different thickness within these subregions, especially in the case of those on the boundary of the HV capacitor. This led to the deterioration of fits since several Gaussian cores of very different widths (corresponding to subregions of different material thickness) were overlayed in the projected angle distributions, preventing the extraction of a reliable value for $\Theta_{x,y}$.

        Combining data from both energies, subregions within the HV hole show an average $x/X_0$ of $(0.365 \pm 0.05)\%$, which is in agreement with the estimated value of $0.394\%$ in \cref{t:estimate}. A contribution of $0.3$-$0.4\%$ from the module rails was observed from approximately row 105 upward, in agreement with the estimated difference of $0.302\%$ for regions with rails. 
        The combined mean of $(0.72 \pm 0.05)\%$ matches well the estimated value of $0.753\%$ for the ROC2 LIN region.

        For both energies, the values of $x/X_0$ are gradually increasing toward the edges of the map in regions that should be of uniform composition. Despite the countermeasures implemented and discussed in \cref{s:edge_region_correction}, residual edge effects appear to remain for subregions far from the centre of the beam spot. These effects are difficult to quantify explicitly since non-uniformity in the material content of the HDI within the accessible region of this ROC prevents the comparison of regions of equal material content at the edges and centre of the beam spot. Consequently, this effect has not been included as part of the quoted systematic uncertainties for the measurement.

        The combined fits for ROC0 at \SI{40}{\MeV} are shown in \cref{f:roc0_40_c}, and the visible features match the X-ray layouts well, with the SMD components and e-link connector clearly visible. However, the residual edge effects appear much stronger, which we attribute to a narrower beam width during the ROC0 \SI{40}{\MeV} data-taking causing a widening of the band of angles suppressed. Despite this, the global mean of $0.95\pm0.09$\% remains in agreement with the ROC0 LIN estimated average of $0.892$\%.
    
        \begin{figure}[t]
            \centering
            \subfigure[ROC2, 40+\SI{65}{\MeV}, data in App. B, \cref{t:roc2map}.]{\label{f:roc2map}%
                \includegraphics[width=.48\textwidth]{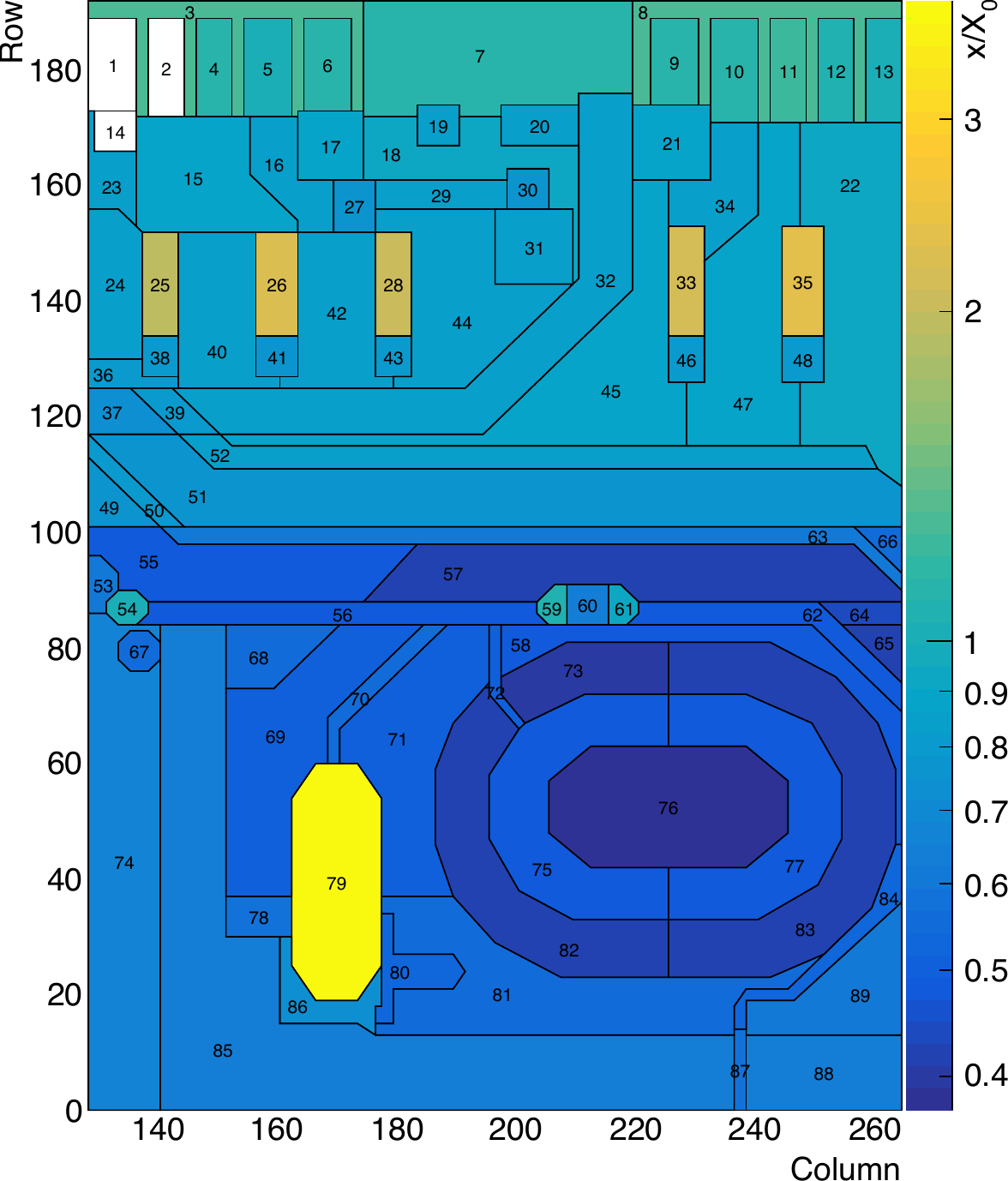}}\hfill%
            \subfigure[ROC0, \SI{40}{\MeV} only, data in App. B, \cref{t:roc0map}.]{\label{f:roc0map}%
                \includegraphics[width=.48\textwidth]{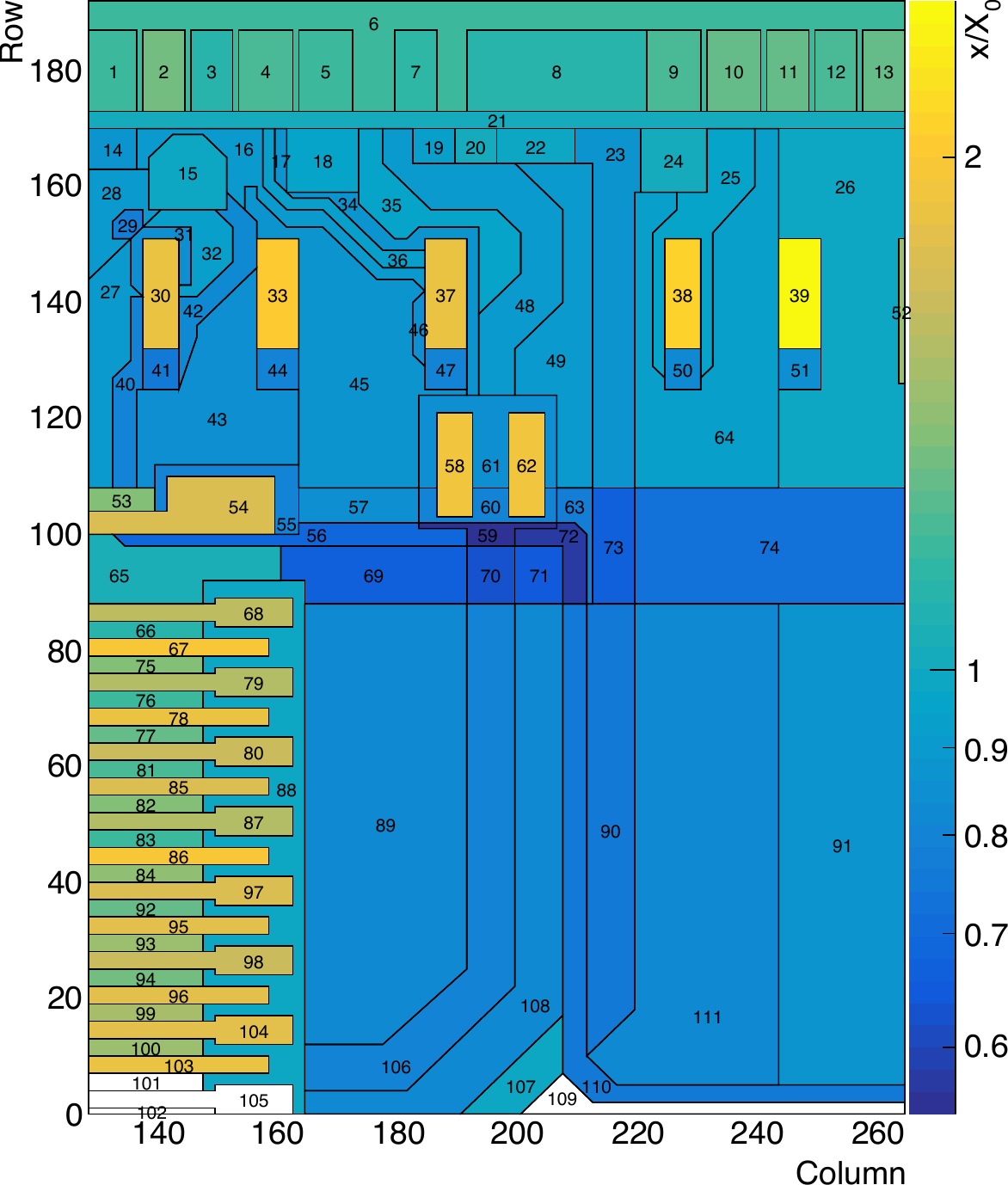}}
            \caption{\label{f:radlen_map}%
                Fractional radiation length maps for polygonal regions of uniform material on the RD53A prototype module derived from GammaPlus fits of the global angle distributions in these regions. The numeric labels in each subregion identify these in the data tables in \cref{s:map_tables}. Empty subregions were not fitted due to insufficient statistics.%
            }
        \end{figure}

        The beam placement and size for the ROC0 \SI{65}{\MeV} runs was later determined to have been quite poor, and this was reflected by the quality of the AGDSCB fits and clear inflation of the fractional radiation length values obtained using all fitting methods toward the edges of the measured region, as shown in the combined value histogram provided in \cref{s:additional_datasets}. 
        Due to these issues, the data recorded with a beam energy of \SI{65}{\MeV} was not used for a combined measurement map in \cref{s:uniform_maps} and instead only the \SI{40}{\MeV} dataset was considered. Results for this dataset and further results for the ROC2 \SI{65}{\MeV} dataset are included in \cref{s:additional_datasets}.

    \end{asection}

    \begin{asection}{Fractional radiation length by region of uniform material composition}%
        \label{s:uniform_maps}

        The aforementioned issues with boundaries between regions of different material composition in rectangular subregions motivated maps with polygonal subregions of approximately uniform material content as shown in \cref{f:radlen_map}. For each subregion, all events passing through pixels with the geometric centre of the pixel enclosed within the polygonal subregion were considered in the fit. The global angle distribution was determined to cope best with the non-rectangular nature of the subregions as opposed to the projected distributions, which deteriorated signficantly in quality for very narrow subregions. This is assumed to be due to the inherent directionality of the projected distributions, which respond to hard material boundaries perpendicular to the projection axis, whilst the global angle is invariant under a rotation about the beam axis, and appears to hence be less susceptible to directional effects. The minimum statistics per subregion was reduced to 4000.
        
        \begin{table}[b]
            \small
            \centering \renewcommand{\arraystretch}{1.2}
            \begin{tabular}{|r||c|ccc|ccc|}
                \hline
                \multirow{2}{*}{Region\hspace{1.75em}} & Estimated & \multicolumn{3}{c|}{ROC2} & \multicolumn{3}{c|}{ROC0} \\
                   & $x/X_0$ (\%) & Subr. & $x/X_0$ (\%) & Dev. ($\sigma$) & Subr. & $x/X_0$ (\%) & Dev. ($\sigma$) \\
                \hline
                Sensor + ROC & 0.394 & 76 & $0.37 \pm 0.06$ & -0.40 & - & - & - \\ 
                \ldots\, + HDI\hspace{0.31em} & 0.553 & 81 & $0.56 \pm 0.08$ & +0.09 & 70 & $0.63 \pm 0.11$ & +0.70 \\ 
                \ldots\, + Rails\hspace{0.020em} & 0.856 & 52 & $0.85 \pm 0.09$ & -0.07 & 49 & $0.90 \pm 0.13$ & +0.34 \\
                PCB Trace\hspace{0.75em} & 0.049--0.056 & 70, 71 & $0.04 \pm 0.07$ & -0.13 & 48, 35 & $0.06 \pm 0.14$ & +0.03 \\
                \hline
            \end{tabular}
            \caption{\label{t:compare_estimates}%
                Comparison of estimated fractional radiation length to measurements for selected sample subregions on ROC0 and ROC2. Sample subregions were chosen close to the centre of the beam spot to limit bias from residual edge effects.%
            }
        \end{table}

        A comparison of key subregions to the estimates in \cref{s:estimate} is provided in \cref{t:compare_estimates}. For both ROCs, measured and estimated values agree within uncertainties for subregions selected as representative of the Sensor + ROC, HDI, and module rails. These subregions were chosen close to the centre of the beam spot to reduce the impact of residual edge effects, and in areas where the X-ray map showed no SMD components or other obstructions.

        In order to investigate the local precision of the measurement, neighbouring subregions differing only by a single trace layer in the HDI were selected for each ROC and compared to the estimated single trace contribution of 0.049--0.056\%. In both cases the measurement matched the trace thickness to 13\% of the standard error or less, potentially indicating substantially better local precision than would be assumed from the uncertainties estimated in this study. This behaviour is also observed for the central regions of ROC0, indicating that the uncertainty due to multiple scattering in air may be significantly less than estimated toward the centre of the beamspot. In general, the uncertainty from multiple scattering in air derived above appears to be overestimated when compared to the local variability between regions of similar material content, and an improved treatment of this uncertainty would be desirable. Deriving an improved uncertainty treatment would likely have required either reference measurements of objects of known fractional radiation length during data-taking, or dedicated simulations of scattering behaviour within the telescope, neither of which were within the scope of this work.

    \end{asection}

\end{asection}

\begin{asection}{Conclusion}%
    \label{s:conclusion}

    A technique has been developed to quantify the material content of a hybrid pixel module via measurement of the multiple scattering of an \si{\MeV}-range positron beam in a three-plane telescope. Measured values of the fractional radiation length for key regions agree with estimates based on the known composition of the module. Combining the 40 and \SI{65}{\MeV} datasets for ROC2 gives an average value of the fractional radiation length of $(0.72 \pm 0.05)\%$ across the surface, and $(0.95 \pm 0.09)\%$ is obtained for ROC0 using only the \SI{40}{\MeV} runs. These measurements match the dedicated predictions for each region to within 4.4\% and 6.5\%, respectively. Differences between the two ROC front-ends may be traced back to the material composition of each region, in particular the SMD connector on ROC0 and the lower density of copper traces in the section of the HDI visible within the ROC2 region.

    Alongside these results, a statistical method for the detection and correction of trigger desynchronisation in multi-plane telescopes utilising inter-plane correlations was developed, and was able to recover 96\% of all runs in which one of the MALTA planes desynchronised, leading to an 18.8\% increase in total event yield. An analysis strategy utilising different fitting methods on the projected and global deflection angles was devised, and corrections due to the geometric limits of the sensor geometry and edge effects due to beam shape were implemented.

    \begin{asection}{Limitations of this technique and proposed improvements for future measurements}

        Several limitations were observed in the measurements, the most influential being residual edge effects persisting despite the introduction of the corrections described in \cref{s:edge_region_correction}. The need for both types of corrections introduced in \cref{s:active_geometry_corrections,s:edge_region_correction} could be mitigated by adding a 2D translation stage to allow the adjustment of the position of the central plane between runs.

        Multiple scattering in air and the beam energy spread dominated the uncertainties in most subregions, indicating that the precision of similar measurements will likely be limited by the telescope and beam properties rather than the achievable statistics. The strongest increase in precision is expected to be achievable through a reduction of the width of the momentum band of the beam, of the inter-plane spacing within the telescope, or of the fractional radiation length of the inter-plane medium\footnote{For example, using a low-Z gas such as helium rather than air: $X_{0,\mathrm{He}} \, (\si{cm}) \approx 18 \cdot X_{0,\mathrm{air}} \, (\si{\cm})$ under laboratory conditions.}.

    \end{asection}
    
    \begin{asection}{Outlook}

        The measurements presented bode well for the design of the CMS RD53A quad modules investigated. Measured values of the fractional radiation length match well with estimates used during the design phase of the new modules, and highlight areas which could be targeted for further reductions. In particular, the module rails contribute a substantial proportion of the total material budget at 17.2\% once averaged over the entire module surface, and a new technique to allow the removal of the rails is currently being investigated. Overall, these modules make a key contribution toward ensuring the material budget target of an $x/X_0$ of 2.5\% for the CMS Phase-2 pixel layers is met~\cite{migliore2016}, and toward the overall goals of maximising vertex resolution and minimising detector material to ensure the detector is ready for the unprecedented operating environment of the HL-LHC\@.

    \end{asection}

\end{asection}

\begin{asection}{Acknowledgements}

    We would like to acknowledge the generous contributions of Heinz Pernegger, Carlos Solans, Valerio Dao, and the ATLAS MALTA group in providing hardware, software, and support for the telescope used for this measurement. The results shown in this paper were collected at the PSI PiE1 facility in Villigen, Switzerland, and we would like to thank 
    the beam physicists and staff at PSI\@.
   
    The tracker groups gratefully acknowledge financial support from the following funding agencies: BMWFW and FWF (Austria); FNRS and FWO (Belgium); CERN\@; MSE and CSF (Croatia); Academy of Finland, MEC, and HIP (Finland); CEA and CNRS/IN2P3 (France); BMBF, DFG, and HGF (Germany); GSRT (Greece); NKFIH K143477 and VLAB at HUN-REN Wigner RCP (Hungary); DAE and DST (India); INFN (Italy); PAEC (Pakistan); SEIDI, CPAN, PCTI and FEDER (Spain); Swiss Funding Agencies (Switzerland); MST (Taipei); STFC (United Kingdom); DOE and NSF (U.S.A.)\@. This project has received funding from the European Union’s Horizon 2020 research and innovation programme under the Marie Sk\l odowska-Curie grant agreement No 884104 (PSI-FELLOW-III-3i)\@. Individuals have received support from HFRI (Greece)\@.

\end{asection}

\bibliography{build/merged}

\vfill

\pagebreak

\appendix

\begin{asection}{\texorpdfstring{\Molex{}}{Molex® }ZIF connector and flex cable material budget estimates}%
    \label{s:molex_estimate}

    The fractional radiation length estimate for the \Molex{}33-pin ZIF connector and flex cable used within the TBPX prototype module HDI design is provided in \cref{t:molex_rl}. The estimates were derived from manufacturer data by smearing volumetric estimates of the material content across the cross-sectional area.

    \begin{table}[h]
        \footnotesize%
        \caption{\label{t:molex_rl}%
            Material budget estimate for the \Molex{}33-pin \SI{0.3}{\mm}-pitch ZIF connector and flex cable, based on design files and data sheets provided in Refs.~\cite{molex_zif,molex_flex}.%
        }%
        \begin{center}%
            \begin{tabular}{|l||cccc|c|}%
                \hline%
                Component & Material & $X_0$ (\si{\cm}) & Est. $h\times w\times d$ (\si{\mm}) & $V$ (\si{\mm^{3}}) & $x/X_0$ (\%) \\
                \hline\hline
                Housing & Liquid Crystal Polymer & 29 & $0.40\times12\times3.45$ & 16.56 & 0.138  \\
                2$\times$ Tails & Phosphor Bronze & 1.41 & $0.15\times(1\times0.8 + 0.5\times2.5)$ & 0.233 & 0.080 \\
                33$\times$ Nails & Phosphor Bronze & 1.41 & $0.1\times0.1\times2.6$ & 0.0374 & 0.208 \\
                33$\times$ Terminals & Phosphor Bronze & 1.41 & $0.1\times0.5\times0.4$ & 0.024 & 0.135 \\
                Cable contacts & Cu, 67\% layer density & 1.436 & $0.035\times2.5\times10.2$ & 0.598 & 0.010 \\
                \hline
                \bf \Molex ZIF conn.  &\bf - &\bf - &\bf $1.15\times12\times3.45$ &\bf - &\bf 0.571 \\
                \hline
                \multicolumn{6}{c}{}\\
                \hline
                Component & Material & $X_0$ (\si{\cm}) & Est. $h$ (\si{\mm}) &   & $x/X_0$ (\%) \\
                \hline\hline
                Coverlay & PI & 29 & $0.025$ & & 0.0086  \\
                Traces & Cu, 50\% trace density & 1.436 & $0.018$ & & 0.0627 \\
                Glues & Acrylic & 42.6 & $0.025$ & & 0.0059 \\
                \hline
                \bf \Molex{} flex  &\bf - &\bf - &\bf 0.068 &\bf &\bf 0.0772 \\
                \hline
            \end{tabular}%
        \end{center}%
    \end{table}

\end{asection}

\clearpage
\begin{asection}{Additional rectangular-binned datasets}%
    \label{s:additional_datasets}

    Additional rectangular-binned combined measurement maps are provided here for both ROC0 and ROC2 to complement the data and comparisons presented in \cref{s:rectangle_results}.

    \begin{figure}[H]
        \centering
        \includegraphics[width=.65\textwidth,page=1]{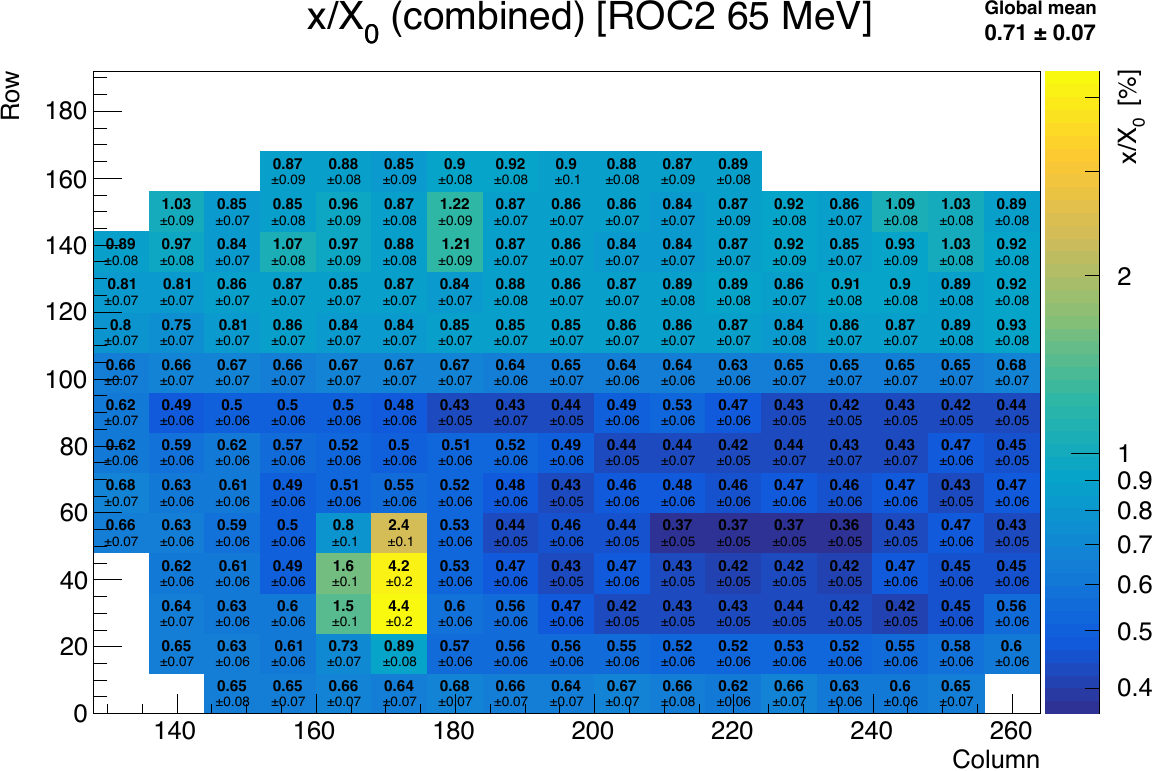}
        \caption{\label{f:roc2_65_c}%
            Combined value fractional radiation length map of ROC2 at \SI{65}{\MeV}. Empty subregions were not fitted due to insufficient statistics.
        }
    \end{figure}
    \begin{figure}[H]
        \centering
        \includegraphics[width=.65\textwidth,page=1]{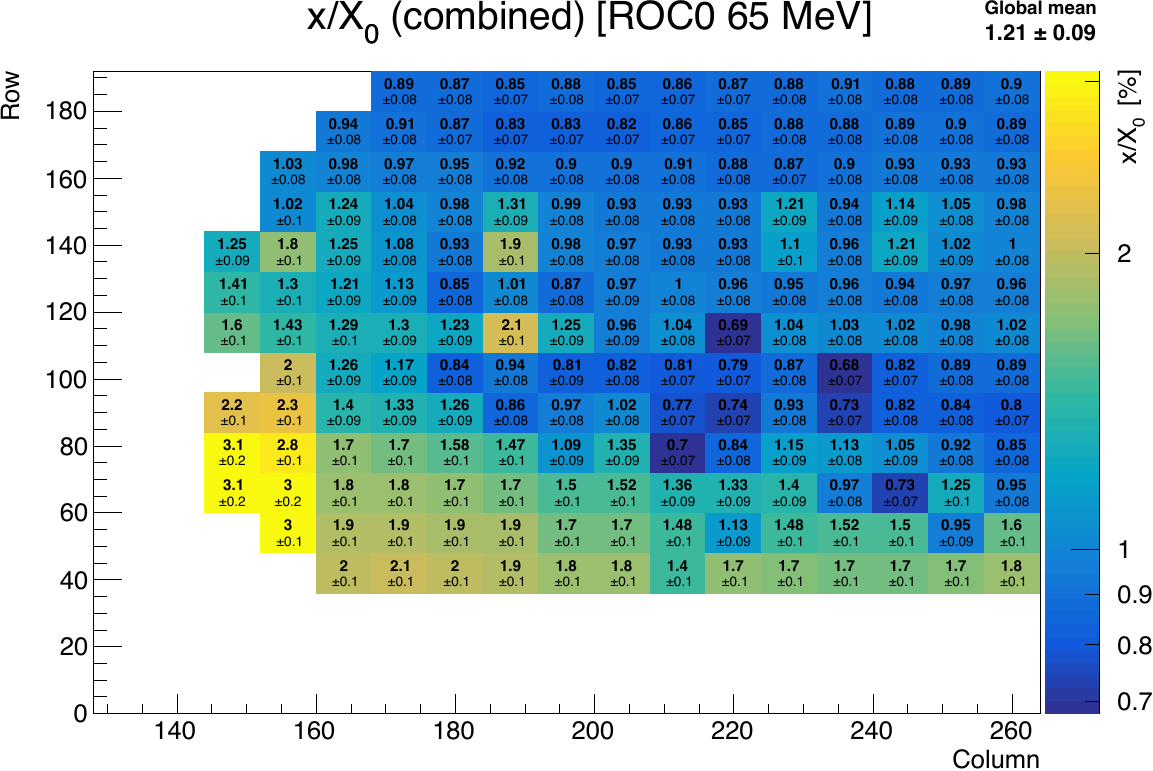}
        \caption{\label{f:roc0_65_c}%
            Combined value fractional radiation length map of ROC0 at \SI{65}{\MeV}. Empty subregions were not fitted due to insufficient statistics. Misalignment of the beam spot during these runs resulted in strong edge-bias effects across the entire dataset, and the data were not used in the final estimates.
        }
    \end{figure}
\end{asection}

\clearpage
\begin{asection}{Data tables for regions of uniform composition}%
    \label{s:map_tables}

    This appendix provides data tables to complement the polygonal maps in \cref{s:uniform_maps}, and is designed to facilitate the easy look-up of fractional radiation length values for key regions of the RD53A prototype module, as given by a global angle fit using the GammaPlus method.

    {\footnotesize\renewcommand{\arraystretch}{1.1}
    \setlength{\LTcapwidth}{\textwidth}
    \begin{longtable}{c || c c c | c c c | c c }
        \caption[]{\label{t:roc2map}Table of numeric values for ROC2 polygonal subregions to complement \cref{f:roc2map}. For each polygonal subregion, combined and individual values of $x/X_0$ are tabulated for the 40 and \SI{65}{\MeV} datasets, and the deviation between the two is given in standard errors. Additionally, the fit quality ($\chi^2$/ndof) and statistics in each subregion are provided.} \\
        \hline\hline
        \multirow{2}{*}{Subregion} & \multicolumn{3}{c|}{\SI{40}{\MeV} GammaPlus fit} & \multicolumn{3}{c|}{\SI{65}{\MeV} GammaPlus fit} & \multicolumn{2}{c}{Combined $40/\SI{65}{\MeV}$} \\
           & $x/X_0$ (\%) & $\chi^2$/ndf & Statistics & $x/X_0$ (\%) & $\chi^2$/ndf & Statistics & $x/X_0$ (\%) & Dev. ($\sigma$) \\
        \hline \endfirsthead
        \caption[]{(continued)} \\
        \hline
        \multirow{2}{*}{Subregion} & \multicolumn{3}{c|}{\SI{40}{\MeV} GammaPlus fit} & \multicolumn{3}{c|}{\SI{65}{\MeV} GammaPlus fit} & \multicolumn{2}{c}{Combined $40/\SI{65}{\MeV}$} \\
           & $x/X_0$ (\%) & $\chi^2$/ndf & Statistics & $x/X_0$ (\%) & $\chi^2$/ndf & Statistics & $x/X_0$ (\%) & Dev. ($\sigma$) \\
        \hline \endhead
        \hline \endfoot
        \hline\hline \endlastfoot
        \input{plots/results/roc2_v5_reorder.txt}
    \end{longtable} }
    \clearpage
    {\footnotesize\renewcommand{\arraystretch}{1.1}
    \setlength{\LTcapwidth}{\textwidth}
    \begin{longtable}{c || c c c | c c c | c c }
        \caption[]{\label{t:roc0map}Table of numeric values for ROC0 polygonal subregions to complement \cref{f:roc0map}. For each polygonal subregion, combined and individual values of $x/X_0$ are tabulated for the 40 and \SI{65}{\MeV} datasets, and the deviation between the two is given in standard errors. Due to the low quality of the \SI{65}{\MeV} dataset, the values shown in \cref{f:roc0map} correspond to the \SI{40}{\MeV} dataset rather than the combined values. Additionally, the fit quality ($\chi^2$/ndof) and statistics in each subregion are provided.} \\
        \hline\hline
        \multirow{2}{*}{Subregion} & \multicolumn{3}{c|}{\SI{40}{\MeV} GammaPlus fit} & \multicolumn{3}{c|}{\SI{65}{\MeV} GammaPlus fit} & \multicolumn{2}{c}{Combined $40/\SI{65}{\MeV}$} \\
         & $x/X_0$ (\%) & $\chi^2$/ndf & Statistics & $x/X_0$ (\%) & $\chi^2$/ndf & Statistics & $x/X_0$ (\%) & Dev. ($\sigma$) \\
        \hline \endfirsthead
        \caption[]{(continued)} \\
        \hline
        \multirow{2}{*}{Subregion} & \multicolumn{3}{c|}{\SI{40}{\MeV} GammaPlus fit} & \multicolumn{3}{c|}{\SI{65}{\MeV} GammaPlus fit} & \multicolumn{2}{c}{Combined $40/\SI{65}{\MeV}$} \\
         & $x/X_0$ (\%) & $\chi^2$/ndf & Statistics & $x/X_0$ (\%) & $\chi^2$/ndf & Statistics & $x/X_0$ (\%) & Dev. ($\sigma$) \\
        \hline \endhead
        \hline \endfoot
        \hline\hline \endlastfoot
        \input{plots/results/roc0_v5.txt}
    \end{longtable} }
\end{asection}

\nolinenumbers%

\include{TrackerAuthorList_2024_Phase2_PixelModule_RadiationLength}

\end{document}

%% file: plots/results/roc2_v5_reorder.txt
\textbf{1} & - & - & $2.6\cdot10^3$ & - & - & $2.6\cdot10^3$ & - & - \\
\textbf{2} & - & - & $3.4\cdot10^3$ & - & - & $3.4\cdot10^3$ & - & - \\
\textbf{3} & $1.29 \pm 0.16$ & $1.06$ & $8.4\cdot10^3$ & - & - & $8.4\cdot10^3$ & $1.29 \pm 0.16$ & - \\
\textbf{4} & $1.09 \pm 0.16$ & $0.8$ & $4.3\cdot10^3$ & - & - & $4.3\cdot10^3$ & $1.09 \pm 0.16$ & - \\
\textbf{5} & $1.01 \pm 0.14$ & $1.38$ & $7.0\cdot10^3$ & - & - & $7.0\cdot10^3$ & $1.01 \pm 0.14$ & - \\
\textbf{6} & $1.12 \pm 0.15$ & $0.8$ & $7.0\cdot10^3$ & - & - & $7.0\cdot10^3$ & $1.12 \pm 0.15$ & - \\
\textbf{7} & $1.13 \pm 0.15$ & $1.2$ & $4.1\cdot10^4$ & - & - & $4.1\cdot10^4$ & $1.13 \pm 0.15$ & - \\
\textbf{8} & $1.27 \pm 0.16$ & $1.48$ & $1.0\cdot10^4$ & - & - & $1.0\cdot10^4$ & $1.27 \pm 0.16$ & - \\
\textbf{9} & $1.10 \pm 0.15$ & $0.97$ & $6.2\cdot10^3$ & - & - & $6.2\cdot10^3$ & $1.10 \pm 0.15$ & - \\
\textbf{10} & $1.09 \pm 0.15$ & $1.41$ & $8.7\cdot10^3$ & - & - & $8.7\cdot10^3$ & $1.09 \pm 0.15$ & - \\
\textbf{11} & $1.20 \pm 0.16$ & $0.85$ & $5.8\cdot10^3$ & - & - & $5.8\cdot10^3$ & $1.20 \pm 0.16$ & - \\
\textbf{12} & $1.05 \pm 0.15$ & $1.6$ & $4.9\cdot10^3$ & - & - & $4.9\cdot10^3$ & $1.05 \pm 0.15$ & - \\
\textbf{13} & $1.03 \pm 0.15$ & $0.98$ & $4.2\cdot10^3$ & - & - & $4.2\cdot10^3$ & $1.03 \pm 0.15$ & - \\
\textbf{14} & - & - & $3.1\cdot10^3$ & - & - & $3.1\cdot10^3$ & - & - \\
\textbf{15} & $0.87 \pm 0.13$ & $0.98$ & $1.1\cdot10^5$ & $0.90 \pm 0.13$ & $1.02$ & $1.1\cdot10^5$ & $0.89 \pm 0.09$ & $0.18$ \\
\textbf{16} & $0.83 \pm 0.13$ & $0.93$ & $5.3\cdot10^4$ & $0.85 \pm 0.13$ & $0.84$ & $5.3\cdot10^4$ & $0.84 \pm 0.09$ & $0.13$ \\
\textbf{17} & $0.86 \pm 0.13$ & $1.0$ & $3.2\cdot10^4$ & $0.94 \pm 0.14$ & $1.24$ & $3.2\cdot10^4$ & $0.90 \pm 0.09$ & $0.42$ \\
\textbf{18} & $0.90 \pm 0.13$ & $0.96$ & $1.1\cdot10^5$ & $0.93 \pm 0.13$ & $0.9$ & $1.1\cdot10^5$ & $0.91 \pm 0.09$ & $0.16$ \\
\textbf{19} & $0.84 \pm 0.13$ & $0.84$ & $9.6\cdot10^3$ & - & - & $9.6\cdot10^3$ & $0.84 \pm 0.13$ & - \\
\textbf{20} & $0.84 \pm 0.13$ & $1.11$ & $1.8\cdot10^4$ & - & - & $1.8\cdot10^4$ & $0.84 \pm 0.13$ & - \\
\textbf{21} & $0.86 \pm 0.13$ & $1.24$ & $4.3\cdot10^4$ & $0.87 \pm 0.14$ & $0.87$ & $4.3\cdot10^4$ & $0.87 \pm 0.09$ & $0.07$ \\
\textbf{22} & $0.92 \pm 0.13$ & $1.06$ & $4.2\cdot10^5$ & $0.92 \pm 0.13$ & $1.14$ & $4.2\cdot10^5$ & $0.92 \pm 0.09$ & $-0.01$ \\
\textbf{23} & $0.85 \pm 0.13$ & $0.94$ & $1.1\cdot10^4$ & - & - & $1.1\cdot10^4$ & $0.85 \pm 0.13$ & - \\
\textbf{24} & $0.87 \pm 0.13$ & $1.36$ & $5.7\cdot10^4$ & $0.84 \pm 0.13$ & $1.15$ & $5.7\cdot10^4$ & $0.86 \pm 0.09$ & $-0.17$ \\
\textbf{25} & $1.97 \pm 0.19$ & $1.37$ & $3.4\cdot10^4$ & $1.98 \pm 0.19$ & $1.28$ & $3.4\cdot10^4$ & $1.97 \pm 0.14$ & $0.04$ \\
\textbf{26} & $2.26 \pm 0.22$ & $0.94$ & $6.3\cdot10^4$ & $2.25 \pm 0.21$ & $0.95$ & $6.3\cdot10^4$ & $2.25 \pm 0.15$ & $-0.02$ \\
\textbf{27} & $0.73 \pm 0.12$ & $1.21$ & $3.3\cdot10^4$ & $0.80 \pm 0.13$ & $0.78$ & $3.3\cdot10^4$ & $0.76 \pm 0.09$ & $0.41$ \\
\textbf{28} & $1.89 \pm 0.19$ & $1.16$ & $7.1\cdot10^4$ & $2.17 \pm 0.20$ & $0.85$ & $7.1\cdot10^4$ & $2.02 \pm 0.14$ & $1.03$ \\
\textbf{29} & $0.81 \pm 0.13$ & $1.12$ & $5.7\cdot10^4$ & $0.85 \pm 0.13$ & $1.59$ & $5.7\cdot10^4$ & $0.83 \pm 0.09$ & $0.2$ \\
\textbf{30} & $0.76 \pm 0.12$ & $0.94$ & $2.5\cdot10^4$ & $0.78 \pm 0.13$ & $0.81$ & $2.5\cdot10^4$ & $0.77 \pm 0.09$ & $0.13$ \\
\textbf{31} & $0.81 \pm 0.13$ & $1.14$ & $1.3\cdot10^5$ & $0.84 \pm 0.13$ & $0.71$ & $1.3\cdot10^5$ & $0.83 \pm 0.09$ & $0.19$ \\
\textbf{32} & $0.82 \pm 0.13$ & $1.21$ & $7.7\cdot10^5$ & $0.86 \pm 0.13$ & $1.08$ & $7.7\cdot10^5$ & $0.84 \pm 0.09$ & $0.18$ \\
\textbf{33} & $2.10 \pm 0.21$ & $1.18$ & $7.4\cdot10^4$ & $2.20 \pm 0.21$ & $0.85$ & $7.4\cdot10^4$ & $2.15 \pm 0.15$ & $0.31$ \\
\textbf{34} & $0.86 \pm 0.13$ & $1.03$ & $9.2\cdot10^4$ & $0.85 \pm 0.13$ & $0.82$ & $9.2\cdot10^4$ & $0.85 \pm 0.09$ & $-0.09$ \\
\textbf{35} & $2.38 \pm 0.24$ & $0.93$ & $6.5\cdot10^4$ & $2.32 \pm 0.21$ & $1.07$ & $6.5\cdot10^4$ & $2.35 \pm 0.16$ & $-0.2$ \\
\textbf{36} & $0.80 \pm 0.13$ & $0.98$ & $1.9\cdot10^4$ & $0.81 \pm 0.13$ & $1.48$ & $1.9\cdot10^4$ & $0.81 \pm 0.09$ & $0.06$ \\
\textbf{37} & $0.72 \pm 0.12$ & $0.64$ & $3.1\cdot10^4$ & $0.74 \pm 0.12$ & $0.98$ & $3.1\cdot10^4$ & $0.73 \pm 0.09$ & $0.14$ \\
\textbf{38} & $0.79 \pm 0.13$ & $1.36$ & $1.9\cdot10^4$ & $0.78 \pm 0.12$ & $0.75$ & $1.9\cdot10^4$ & $0.79 \pm 0.09$ & $-0.06$ \\
\textbf{39} & $0.76 \pm 0.12$ & $1.07$ & $3.0\cdot10^4$ & $0.83 \pm 0.13$ & $0.79$ & $3.0\cdot10^4$ & $0.80 \pm 0.09$ & $0.4$ \\
\textbf{40} & $0.83 \pm 0.13$ & $1.15$ & $2.0\cdot10^5$ & $0.85 \pm 0.13$ & $1.3$ & $2.0\cdot10^5$ & $0.84 \pm 0.09$ & $0.16$ \\
\textbf{41} & $0.75 \pm 0.12$ & $0.87$ & $4.0\cdot10^4$ & $0.80 \pm 0.13$ & $1.04$ & $4.0\cdot10^4$ & $0.77 \pm 0.09$ & $0.3$ \\
\textbf{42} & $0.83 \pm 0.13$ & $0.84$ & $3.0\cdot10^5$ & $0.87 \pm 0.13$ & $1.18$ & $3.0\cdot10^5$ & $0.85 \pm 0.09$ & $0.23$ \\
\textbf{43} & $0.77 \pm 0.12$ & $1.2$ & $4.4\cdot10^4$ & $0.82 \pm 0.13$ & $1.0$ & $4.4\cdot10^4$ & $0.79 \pm 0.09$ & $0.26$ \\
\textbf{44} & $0.84 \pm 0.13$ & $1.46$ & $5.1\cdot10^5$ & $0.87 \pm 0.13$ & $1.0$ & $5.1\cdot10^5$ & $0.85 \pm 0.09$ & $0.16$ \\
\textbf{45} & $0.88 \pm 0.13$ & $1.03$ & $7.8\cdot10^5$ & $0.90 \pm 0.13$ & $1.27$ & $7.8\cdot10^5$ & $0.89 \pm 0.09$ & $0.11$ \\
\textbf{46} & $0.79 \pm 0.12$ & $0.75$ & $5.0\cdot10^4$ & $0.82 \pm 0.13$ & $1.06$ & $5.0\cdot10^4$ & $0.81 \pm 0.09$ & $0.22$ \\
\textbf{47} & $0.89 \pm 0.13$ & $0.95$ & $5.1\cdot10^5$ & $0.90 \pm 0.13$ & $0.85$ & $5.1\cdot10^5$ & $0.89 \pm 0.09$ & $0.03$ \\
\textbf{48} & $0.81 \pm 0.13$ & $1.0$ & $4.5\cdot10^4$ & $0.82 \pm 0.13$ & $0.88$ & $4.5\cdot10^4$ & $0.81 \pm 0.09$ & $0.07$ \\
\textbf{49} & $0.78 \pm 0.12$ & $1.37$ & $2.3\cdot10^4$ & $0.75 \pm 0.12$ & $1.16$ & $2.3\cdot10^4$ & $0.76 \pm 0.09$ & $-0.13$ \\
\textbf{50} & $0.82 \pm 0.13$ & $0.88$ & $2.1\cdot10^4$ & $0.82 \pm 0.13$ & $1.35$ & $2.1\cdot10^4$ & $0.82 \pm 0.09$ & $0.0$ \\
\textbf{51} & $0.75 \pm 0.12$ & $0.79$ & $1.3\cdot10^6$ & $0.77 \pm 0.12$ & $1.16$ & $1.3\cdot10^6$ & $0.76 \pm 0.09$ & $0.1$ \\
\textbf{52} & $0.84 \pm 0.13$ & $1.07$ & $4.6\cdot10^5$ & $0.86 \pm 0.13$ & $1.12$ & $4.6\cdot10^5$ & $0.85 \pm 0.09$ & $0.12$ \\
\textbf{53} & $0.63 \pm 0.11$ & $1.33$ & $1.1\cdot10^4$ & $0.58 \pm 0.11$ & $1.03$ & $1.1\cdot10^4$ & $0.60 \pm 0.08$ & $-0.33$ \\
\textbf{54} & $0.99 \pm 0.14$ & $0.94$ & $1.2\cdot10^4$ & $1.09 \pm 0.15$ & $0.94$ & $1.2\cdot10^4$ & $1.03 \pm 0.10$ & $0.48$ \\
\textbf{55} & $0.48 \pm 0.10$ & $1.13$ & $3.8\cdot10^5$ & $0.48 \pm 0.10$ & $1.54$ & $3.8\cdot10^5$ & $0.48 \pm 0.07$ & $0.01$ \\
\textbf{56} & $0.48 \pm 0.10$ & $1.56$ & $2.7\cdot10^5$ & $0.49 \pm 0.10$ & $1.1$ & $2.7\cdot10^5$ & $0.49 \pm 0.07$ & $0.06$ \\
\textbf{57} & $0.41 \pm 0.09$ & $1.61$ & $9.1\cdot10^5$ & $0.41 \pm 0.09$ & $1.87$ & $9.1\cdot10^5$ & $0.41 \pm 0.07$ & $-0.03$ \\
\textbf{58} & $0.49 \pm 0.10$ & $1.45$ & $3.4\cdot10^5$ & $0.49 \pm 0.10$ & $1.18$ & $3.4\cdot10^5$ & $0.49 \pm 0.07$ & $-0.04$ \\
\textbf{59} & $1.07 \pm 0.14$ & $1.01$ & $3.2\cdot10^4$ & $1.08 \pm 0.15$ & $1.0$ & $3.2\cdot10^4$ & $1.07 \pm 0.10$ & $0.05$ \\
\textbf{60} & $0.65 \pm 0.11$ & $1.03$ & $6.4\cdot10^4$ & $0.65 \pm 0.11$ & $1.05$ & $6.4\cdot10^4$ & $0.65 \pm 0.08$ & $0.03$ \\
\textbf{61} & $0.94 \pm 0.14$ & $1.1$ & $3.7\cdot10^4$ & $0.93 \pm 0.13$ & $1.1$ & $3.7\cdot10^4$ & $0.93 \pm 0.10$ & $-0.08$ \\
\textbf{62} & $0.48 \pm 0.10$ & $1.37$ & $1.9\cdot10^5$ & $0.47 \pm 0.10$ & $1.18$ & $1.9\cdot10^5$ & $0.47 \pm 0.07$ & $-0.04$ \\
\textbf{63} & $0.60 \pm 0.11$ & $1.41$ & $3.8\cdot10^5$ & $0.60 \pm 0.11$ & $1.64$ & $3.8\cdot10^5$ & $0.60 \pm 0.08$ & $0.04$ \\
\textbf{64} & $0.45 \pm 0.10$ & $1.31$ & $3.5\cdot10^4$ & $0.44 \pm 0.10$ & $1.05$ & $3.5\cdot10^4$ & $0.44 \pm 0.07$ & $-0.12$ \\
\textbf{65} & $0.43 \pm 0.09$ & $0.98$ & $3.3\cdot10^4$ & $0.40 \pm 0.09$ & $1.24$ & $3.3\cdot10^4$ & $0.42 \pm 0.07$ & $-0.24$ \\
\textbf{66} & $0.56 \pm 0.11$ & $0.9$ & $1.9\cdot10^4$ & $0.54 \pm 0.11$ & $0.75$ & $1.9\cdot10^4$ & $0.55 \pm 0.08$ & $-0.17$ \\
\textbf{67} & $0.55 \pm 0.11$ & $1.17$ & $1.6\cdot10^4$ & $0.54 \pm 0.11$ & $0.91$ & $1.6\cdot10^4$ & $0.54 \pm 0.07$ & $-0.05$ \\
\textbf{68} & $0.54 \pm 0.10$ & $1.07$ & $1.3\cdot10^5$ & $0.56 \pm 0.11$ & $0.81$ & $1.3\cdot10^5$ & $0.55 \pm 0.07$ & $0.11$ \\
\textbf{69} & $0.49 \pm 0.10$ & $1.29$ & $5.9\cdot10^5$ & $0.51 \pm 0.10$ & $1.26$ & $5.9\cdot10^5$ & $0.50 \pm 0.07$ & $0.12$ \\
\textbf{70} & $0.54 \pm 0.10$ & $1.14$ & $9.5\cdot10^4$ & $0.55 \pm 0.11$ & $1.04$ & $9.5\cdot10^4$ & $0.55 \pm 0.07$ & $0.07$ \\
\textbf{71} & $0.51 \pm 0.10$ & $1.36$ & $7.2\cdot10^5$ & $0.52 \pm 0.10$ & $1.17$ & $7.2\cdot10^5$ & $0.51 \pm 0.07$ & $0.07$ \\
\textbf{72} & $0.50 \pm 0.10$ & $0.99$ & $4.6\cdot10^4$ & $0.51 \pm 0.10$ & $1.12$ & $4.6\cdot10^4$ & $0.50 \pm 0.07$ & $0.01$ \\
\textbf{73} & $0.41 \pm 0.09$ & $1.1$ & $3.4\cdot10^5$ & $0.41 \pm 0.09$ & $1.02$ & $3.4\cdot10^5$ & $0.41 \pm 0.07$ & $0.02$ \\
\textbf{74} & $0.68 \pm 0.12$ & $1.14$ & $2.5\cdot10^5$ & $0.64 \pm 0.11$ & $0.77$ & $2.5\cdot10^5$ & $0.66 \pm 0.08$ & $-0.21$ \\
\textbf{75} & $0.48 \pm 0.10$ & $1.48$ & $8.5\cdot10^5$ & $0.48 \pm 0.10$ & $1.61$ & $8.5\cdot10^5$ & $0.48 \pm 0.07$ & $0.0$ \\
\textbf{76} & $0.37 \pm 0.09$ & $1.71$ & $9.8\cdot10^5$ & $0.37 \pm 0.09$ & $1.51$ & $9.8\cdot10^5$ & $0.37 \pm 0.06$ & $0.01$ \\
\textbf{77} & $0.49 \pm 0.10$ & $1.4$ & $6.6\cdot10^5$ & $0.48 \pm 0.10$ & $1.17$ & $6.6\cdot10^5$ & $0.48 \pm 0.07$ & $-0.02$ \\
\textbf{78} & $0.55 \pm 0.11$ & $1.31$ & $5.3\cdot10^4$ & $0.59 \pm 0.11$ & $1.12$ & $5.3\cdot10^4$ & $0.57 \pm 0.08$ & $0.24$ \\
\textbf{79} & $3.73 \pm 0.28$ & $0.71$ & $2.9\cdot10^5$ & $3.95 \pm 0.27$ & $0.9$ & $2.9\cdot10^5$ & $3.84 \pm 0.20$ & $0.55$ \\
\textbf{80} & $0.54 \pm 0.10$ & $1.09$ & $1.0\cdot10^5$ & $0.52 \pm 0.10$ & $1.25$ & $1.0\cdot10^5$ & $0.53 \pm 0.07$ & $-0.15$ \\
\textbf{81} & $0.57 \pm 0.11$ & $1.55$ & $7.7\cdot10^5$ & $0.55 \pm 0.10$ & $0.87$ & $7.7\cdot10^5$ & $0.56 \pm 0.08$ & $-0.12$ \\
\textbf{82} & $0.42 \pm 0.09$ & $1.39$ & $8.2\cdot10^5$ & $0.42 \pm 0.09$ & $1.23$ & $8.2\cdot10^5$ & $0.42 \pm 0.07$ & $-0.02$ \\
\textbf{83} & $0.42 \pm 0.09$ & $1.66$ & $8.9\cdot10^5$ & $0.42 \pm 0.09$ & $1.51$ & $8.9\cdot10^5$ & $0.42 \pm 0.07$ & $-0.07$ \\
\textbf{84} & $0.53 \pm 0.10$ & $1.26$ & $8.0\cdot10^4$ & $0.51 \pm 0.10$ & $0.92$ & $8.0\cdot10^4$ & $0.52 \pm 0.07$ & $-0.14$ \\
\textbf{85} & $0.65 \pm 0.11$ & $1.6$ & $1.2\cdot10^6$ & $0.64 \pm 0.11$ & $1.28$ & $1.2\cdot10^6$ & $0.64 \pm 0.08$ & $-0.09$ \\
\textbf{86} & $0.72 \pm 0.12$ & $1.09$ & $7.4\cdot10^4$ & $0.73 \pm 0.12$ & $1.1$ & $7.4\cdot10^4$ & $0.73 \pm 0.08$ & $0.03$ \\
\textbf{87} & $0.58 \pm 0.11$ & $1.38$ & $1.8\cdot10^4$ & $0.59 \pm 0.11$ & $1.27$ & $1.8\cdot10^4$ & $0.59 \pm 0.08$ & $0.04$ \\
\textbf{88} & $0.67 \pm 0.11$ & $0.82$ & $1.7\cdot10^5$ & $0.61 \pm 0.11$ & $0.91$ & $1.7\cdot10^5$ & $0.64 \pm 0.08$ & $-0.41$ \\
\textbf{89} & $0.62 \pm 0.11$ & $1.31$ & $2.0\cdot10^5$ & $0.59 \pm 0.11$ & $1.07$ & $2.0\cdot10^5$ & $0.60 \pm 0.08$ & $-0.18$ \\

%% file: plots/results/roc0_v5.txt
\textbf{1} & $1.18 \pm 0.15$ & $1.42$ & $1.2\cdot10^4$ & - & - & $1.2\cdot10^4$ & $1.18 \pm 0.15$ & - \\
\textbf{2} & $1.34 \pm 0.16$ & $1.0$ & $1.5\cdot10^4$ & - & - & $1.5\cdot10^4$ & $1.34 \pm 0.16$ & - \\
\textbf{3} & $1.12 \pm 0.15$ & $1.01$ & $2.0\cdot10^4$ & $1.07 \pm 0.15$ & $1.34$ & $2.0\cdot10^4$ & $1.09 \pm 0.11$ & $-0.24$ \\
\textbf{4} & $1.23 \pm 0.15$ & $1.15$ & $3.2\cdot10^4$ & $1.06 \pm 0.15$ & $1.35$ & $3.2\cdot10^4$ & $1.14 \pm 0.11$ & $-0.81$ \\
\textbf{5} & $1.17 \pm 0.15$ & $1.12$ & $3.7\cdot10^4$ & $0.98 \pm 0.14$ & $1.19$ & $3.7\cdot10^4$ & $1.07 \pm 0.10$ & $-0.9$ \\
\textbf{6} & $1.17 \pm 0.15$ & $1.49$ & $1.8\cdot10^5$ & $0.87 \pm 0.13$ & $1.35$ & $1.8\cdot10^5$ & $1.00 \pm 0.10$ & $-1.53$ \\
\textbf{7} & $1.15 \pm 0.15$ & $1.42$ & $3.1\cdot10^4$ & $0.91 \pm 0.13$ & $1.12$ & $3.1\cdot10^4$ & $1.02 \pm 0.10$ & $-1.22$ \\
\textbf{8} & $1.11 \pm 0.14$ & $1.08$ & $1.1\cdot10^5$ & $0.87 \pm 0.13$ & $1.1$ & $1.1\cdot10^5$ & $0.98 \pm 0.10$ & $-1.21$ \\
\textbf{9} & $1.23 \pm 0.15$ & $0.95$ & $2.7\cdot10^4$ & $0.91 \pm 0.13$ & $1.36$ & $2.7\cdot10^4$ & $1.05 \pm 0.10$ & $-1.6$ \\
\textbf{10} & $1.29 \pm 0.16$ & $1.1$ & $2.3\cdot10^4$ & $0.92 \pm 0.13$ & $1.36$ & $2.3\cdot10^4$ & $1.07 \pm 0.10$ & $-1.78$ \\
\textbf{11} & $1.28 \pm 0.16$ & $1.26$ & $1.5\cdot10^4$ & $0.91 \pm 0.13$ & $1.44$ & $1.5\cdot10^4$ & $1.07 \pm 0.10$ & $-1.78$ \\
\textbf{12} & $1.22 \pm 0.16$ & $1.0$ & $1.2\cdot10^4$ & $0.91 \pm 0.13$ & $0.79$ & $1.2\cdot10^4$ & $1.04 \pm 0.10$ & $-1.52$ \\
\textbf{13} & $1.30 \pm 0.17$ & $0.96$ & $9.9\cdot10^3$ & $0.92 \pm 0.13$ & $1.2$ & $9.9\cdot10^3$ & $1.06 \pm 0.11$ & $-1.77$ \\
\textbf{14} & $0.89 \pm 0.13$ & $1.18$ & $1.0\cdot10^4$ & - & - & $1.0\cdot10^4$ & $0.89 \pm 0.13$ & - \\
\textbf{15} & $0.98 \pm 0.14$ & $0.92$ & $5.3\cdot10^4$ & $1.22 \pm 0.16$ & $0.85$ & $5.3\cdot10^4$ & $1.08 \pm 0.10$ & $1.14$ \\
\textbf{16} & $0.91 \pm 0.13$ & $1.17$ & $9.0\cdot10^4$ & $1.14 \pm 0.15$ & $1.06$ & $9.0\cdot10^4$ & $1.01 \pm 0.10$ & $1.15$ \\
\textbf{17} & $0.89 \pm 0.13$ & $1.04$ & $8.2\cdot10^3$ & - & - & $8.2\cdot10^3$ & $0.89 \pm 0.13$ & - \\
\textbf{18} & $0.97 \pm 0.14$ & $0.96$ & $6.4\cdot10^4$ & $0.96 \pm 0.14$ & $1.21$ & $6.4\cdot10^4$ & $0.96 \pm 0.10$ & $-0.03$ \\
\textbf{19} & $0.94 \pm 0.14$ & $1.1$ & $2.0\cdot10^4$ & $0.86 \pm 0.13$ & $1.0$ & $2.0\cdot10^4$ & $0.89 \pm 0.09$ & $-0.43$ \\
\textbf{20} & $1.01 \pm 0.14$ & $1.01$ & $1.9\cdot10^4$ & $0.91 \pm 0.13$ & $1.05$ & $1.9\cdot10^4$ & $0.95 \pm 0.10$ & $-0.5$ \\
\textbf{21} & $1.02 \pm 0.14$ & $1.03$ & $1.2\cdot10^5$ & $0.86 \pm 0.13$ & $1.07$ & $1.2\cdot10^5$ & $0.94 \pm 0.10$ & $-0.85$ \\
\textbf{22} & $0.99 \pm 0.14$ & $0.93$ & $3.3\cdot10^4$ & $0.85 \pm 0.13$ & $1.0$ & $3.3\cdot10^4$ & $0.91 \pm 0.09$ & $-0.71$ \\
\textbf{23} & $0.88 \pm 0.13$ & $1.29$ & $3.3\cdot10^5$ & $0.93 \pm 0.13$ & $1.33$ & $3.3\cdot10^5$ & $0.90 \pm 0.09$ & $0.3$ \\
\textbf{24} & $1.01 \pm 0.14$ & $0.75$ & $4.2\cdot10^4$ & $0.87 \pm 0.13$ & $0.88$ & $4.2\cdot10^4$ & $0.93 \pm 0.10$ & $-0.73$ \\
\textbf{25} & $0.97 \pm 0.14$ & $1.03$ & $1.2\cdot10^5$ & $0.92 \pm 0.13$ & $1.41$ & $1.2\cdot10^5$ & $0.95 \pm 0.10$ & $-0.24$ \\
\textbf{26} & $0.99 \pm 0.14$ & $0.77$ & $4.7\cdot10^5$ & $0.98 \pm 0.14$ & $2.31$ & $4.7\cdot10^5$ & $0.98 \pm 0.10$ & $-0.03$ \\
\textbf{27} & $0.89 \pm 0.13$ & $1.19$ & $1.1\cdot10^5$ & $1.79 \pm 0.19$ & $1.38$ & $1.1\cdot10^5$ & $1.19 \pm 0.11$ & $3.92$ \\
\textbf{28} & $0.90 \pm 0.13$ & $1.03$ & $3.5\cdot10^4$ & $1.53 \pm 0.19$ & $1.56$ & $3.5\cdot10^4$ & $1.10 \pm 0.11$ & $2.69$ \\
\textbf{29} & $0.78 \pm 0.13$ & $1.22$ & $6.8\cdot10^3$ & - & - & $6.8\cdot10^3$ & $0.78 \pm 0.13$ & - \\
\textbf{30} & $1.88 \pm 0.19$ & $1.03$ & $5.6\cdot10^4$ & $3.41 \pm 0.28$ & $0.98$ & $5.6\cdot10^4$ & $2.36 \pm 0.16$ & $4.55$ \\
\textbf{31} & $0.85 \pm 0.13$ & $1.58$ & $2.2\cdot10^4$ & - & - & $2.2\cdot10^4$ & $0.85 \pm 0.13$ & - \\
\textbf{32} & $0.92 \pm 0.13$ & $0.8$ & $5.7\cdot10^4$ & $1.33 \pm 0.16$ & $0.96$ & $5.7\cdot10^4$ & $1.08 \pm 0.10$ & $1.97$ \\
\textbf{33} & $2.01 \pm 0.19$ & $1.05$ & $8.7\cdot10^4$ & $3.27 \pm 0.30$ & $0.89$ & $8.7\cdot10^4$ & $2.37 \pm 0.16$ & $3.52$ \\
\textbf{34} & $0.87 \pm 0.13$ & $0.9$ & $9.9\cdot10^4$ & $0.98 \pm 0.14$ & $1.05$ & $9.9\cdot10^4$ & $0.92 \pm 0.09$ & $0.55$ \\
\textbf{35} & $0.96 \pm 0.14$ & $0.99$ & $1.6\cdot10^5$ & $1.02 \pm 0.14$ & $0.89$ & $1.6\cdot10^5$ & $0.99 \pm 0.10$ & $0.3$ \\
\textbf{36} & $0.93 \pm 0.13$ & $1.03$ & $5.5\cdot10^4$ & $1.10 \pm 0.15$ & $0.96$ & $5.5\cdot10^4$ & $1.01 \pm 0.10$ & $0.83$ \\
\textbf{37} & $1.84 \pm 0.18$ & $1.07$ & $9.8\cdot10^4$ & $2.41 \pm 0.21$ & $1.18$ & $9.8\cdot10^4$ & $2.08 \pm 0.14$ & $1.99$ \\
\textbf{38} & $2.09 \pm 0.20$ & $1.2$ & $6.1\cdot10^4$ & $2.42 \pm 0.21$ & $0.8$ & $6.1\cdot10^4$ & $2.24 \pm 0.14$ & $1.13$ \\
\textbf{39} & $2.47 \pm 0.26$ & $1.1$ & $5.2\cdot10^4$ & $2.54 \pm 0.22$ & $1.0$ & $5.2\cdot10^4$ & $2.51 \pm 0.17$ & $0.2$ \\
\textbf{40} & $0.80 \pm 0.12$ & $0.97$ & $7.1\cdot10^4$ & $1.60 \pm 0.18$ & $1.0$ & $7.1\cdot10^4$ & $1.07 \pm 0.10$ & $3.69$ \\
\textbf{41} & $0.76 \pm 0.12$ & $1.09$ & $2.9\cdot10^4$ & - & - & $2.9\cdot10^4$ & $0.76 \pm 0.12$ & - \\
\textbf{42} & $0.83 \pm 0.13$ & $1.02$ & $7.9\cdot10^4$ & $1.23 \pm 0.16$ & $1.11$ & $7.9\cdot10^4$ & $0.99 \pm 0.10$ & $2.03$ \\
\textbf{43} & $0.84 \pm 0.13$ & $0.76$ & $4.9\cdot10^5$ & $1.48 \pm 0.17$ & $1.06$ & $4.9\cdot10^5$ & $1.08 \pm 0.10$ & $3.07$ \\
\textbf{44} & $0.79 \pm 0.12$ & $1.14$ & $4.9\cdot10^4$ & $1.17 \pm 0.15$ & $1.03$ & $4.9\cdot10^4$ & $0.94 \pm 0.10$ & $1.94$ \\
\textbf{45} & $0.87 \pm 0.13$ & $1.03$ & $8.8\cdot10^5$ & $1.22 \pm 0.15$ & $1.13$ & $8.8\cdot10^5$ & $1.02 \pm 0.10$ & $1.78$ \\
\textbf{46} & $0.84 \pm 0.13$ & $1.39$ & $1.9\cdot10^4$ & $0.96 \pm 0.14$ & $1.44$ & $1.9\cdot10^4$ & $0.89 \pm 0.09$ & $0.63$ \\
\textbf{47} & $0.80 \pm 0.12$ & $1.08$ & $5.3\cdot10^4$ & $0.98 \pm 0.14$ & $1.19$ & $5.3\cdot10^4$ & $0.88 \pm 0.09$ & $0.97$ \\
\textbf{48} & $0.90 \pm 0.13$ & $1.28$ & $3.2\cdot10^5$ & $0.96 \pm 0.14$ & $0.88$ & $3.2\cdot10^5$ & $0.93 \pm 0.09$ & $0.36$ \\
\textbf{49} & $0.90 \pm 0.13$ & $1.61$ & $3.5\cdot10^5$ & $1.01 \pm 0.14$ & $1.05$ & $3.5\cdot10^5$ & $0.95 \pm 0.10$ & $0.59$ \\
\textbf{50} & $0.82 \pm 0.13$ & $1.01$ & $3.4\cdot10^4$ & $0.90 \pm 0.13$ & $1.08$ & $3.4\cdot10^4$ & $0.86 \pm 0.09$ & $0.41$ \\
\textbf{51} & $0.84 \pm 0.13$ & $1.03$ & $3.0\cdot10^4$ & $0.90 \pm 0.13$ & $1.12$ & $3.0\cdot10^4$ & $0.87 \pm 0.09$ & $0.32$ \\
\textbf{52} & $1.46 \pm 0.17$ & $1.06$ & $9.2\cdot10^3$ & $1.63 \pm 0.18$ & $1.01$ & $9.2\cdot10^3$ & $1.54 \pm 0.12$ & $0.66$ \\
\textbf{53} & $1.36 \pm 0.16$ & $0.89$ & $2.9\cdot10^4$ & - & - & $2.9\cdot10^4$ & $1.36 \pm 0.16$ & - \\
\textbf{54} & $1.73 \pm 0.18$ & $1.07$ & $1.9\cdot10^5$ & $3.59 \pm 0.28$ & $1.18$ & $1.9\cdot10^5$ & $2.28 \pm 0.15$ & $5.58$ \\
\textbf{55} & $0.81 \pm 0.13$ & $0.98$ & $1.2\cdot10^5$ & $1.49 \pm 0.17$ & $1.13$ & $1.2\cdot10^5$ & $1.05 \pm 0.10$ & $3.23$ \\
\textbf{56} & $0.68 \pm 0.12$ & $0.98$ & $2.3\cdot10^5$ & $1.27 \pm 0.16$ & $0.76$ & $2.3\cdot10^5$ & $0.89 \pm 0.09$ & $3.02$ \\
\textbf{57} & $0.84 \pm 0.13$ & $1.06$ & $1.7\cdot10^5$ & $1.45 \pm 0.17$ & $1.35$ & $1.7\cdot10^5$ & $1.07 \pm 0.10$ & $2.91$ \\
\textbf{58} & $1.89 \pm 0.19$ & $1.06$ & $1.2\cdot10^5$ & $2.87 \pm 0.24$ & $0.77$ & $1.2\cdot10^5$ & $2.26 \pm 0.15$ & $3.23$ \\
\textbf{59} & $0.55 \pm 0.10$ & $1.24$ & $7.3\cdot10^4$ & $0.83 \pm 0.13$ & $0.95$ & $7.3\cdot10^4$ & $0.66 \pm 0.08$ & $1.73$ \\
\textbf{60} & $0.81 \pm 0.13$ & $0.85$ & $1.1\cdot10^5$ & $1.19 \pm 0.15$ & $0.77$ & $1.1\cdot10^5$ & $0.96 \pm 0.10$ & $1.92$ \\
\textbf{61} & $0.85 \pm 0.13$ & $0.67$ & $2.6\cdot10^5$ & $1.10 \pm 0.14$ & $1.14$ & $2.6\cdot10^5$ & $0.96 \pm 0.10$ & $1.26$ \\
\textbf{62} & $1.92 \pm 0.19$ & $0.84$ & $1.1\cdot10^5$ & $2.77 \pm 0.23$ & $1.6$ & $1.1\cdot10^5$ & $2.26 \pm 0.15$ & $2.81$ \\
\textbf{63} & $0.78 \pm 0.12$ & $1.24$ & $6.9\cdot10^4$ & $1.08 \pm 0.14$ & $1.05$ & $6.9\cdot10^4$ & $0.91 \pm 0.09$ & $1.55$ \\
\textbf{64} & $0.93 \pm 0.13$ & $1.5$ & $6.6\cdot10^5$ & $0.99 \pm 0.14$ & $1.22$ & $6.6\cdot10^5$ & $0.95 \pm 0.10$ & $0.31$ \\
\textbf{65} & $1.03 \pm 0.14$ & $1.15$ & $2.9\cdot10^5$ & $2.25 \pm 0.21$ & $0.94$ & $2.9\cdot10^5$ & $1.42 \pm 0.12$ & $4.91$ \\
\textbf{66} & $1.11 \pm 0.15$ & $0.92$ & $5.5\cdot10^4$ & $2.56 \pm 0.24$ & $1.45$ & $5.5\cdot10^4$ & $1.51 \pm 0.12$ & $5.23$ \\
\textbf{67} & $1.95 \pm 0.20$ & $0.94$ & $8.6\cdot10^4$ & $4.08 \pm 0.32$ & $1.0$ & $8.6\cdot10^4$ & $2.54 \pm 0.17$ & $5.64$ \\
\textbf{68} & $1.61 \pm 0.17$ & $0.98$ & $1.3\cdot10^5$ & $3.68 \pm 0.33$ & $0.87$ & $1.3\cdot10^5$ & $2.06 \pm 0.15$ & $5.56$ \\
\textbf{69} & $0.67 \pm 0.11$ & $0.84$ & $4.8\cdot10^5$ & $1.23 \pm 0.15$ & $0.97$ & $4.8\cdot10^5$ & $0.87 \pm 0.09$ & $2.91$ \\
\textbf{70} & $0.63 \pm 0.11$ & $1.26$ & $1.3\cdot10^5$ & $1.02 \pm 0.14$ & $0.75$ & $1.3\cdot10^5$ & $0.78 \pm 0.09$ & $2.17$ \\
\textbf{71} & $0.65 \pm 0.11$ & $1.13$ & $1.3\cdot10^5$ & $0.99 \pm 0.14$ & $1.18$ & $1.3\cdot10^5$ & $0.79 \pm 0.09$ & $1.92$ \\
\textbf{72} & $0.58 \pm 0.11$ & $0.95$ & $1.2\cdot10^5$ & $0.88 \pm 0.13$ & $1.02$ & $1.2\cdot10^5$ & $0.70 \pm 0.08$ & $1.8$ \\
\textbf{73} & $0.66 \pm 0.11$ & $1.03$ & $2.0\cdot10^5$ & $0.94 \pm 0.13$ & $1.1$ & $2.0\cdot10^5$ & $0.78 \pm 0.09$ & $1.55$ \\
\textbf{74} & $0.74 \pm 0.12$ & $1.55$ & $9.8\cdot10^5$ & $0.94 \pm 0.13$ & $1.89$ & $9.8\cdot10^5$ & $0.83 \pm 0.09$ & $1.12$ \\
\textbf{75} & $1.37 \pm 0.16$ & $1.1$ & $5.4\cdot10^4$ & $2.87 \pm 0.25$ & $1.25$ & $5.4\cdot10^4$ & $1.83 \pm 0.14$ & $5.06$ \\
\textbf{76} & $1.16 \pm 0.15$ & $1.17$ & $5.7\cdot10^4$ & $2.51 \pm 0.23$ & $0.99$ & $5.7\cdot10^4$ & $1.57 \pm 0.12$ & $4.93$ \\
\textbf{77} & $1.30 \pm 0.16$ & $1.04$ & $5.4\cdot10^4$ & $3.3 \pm 0.4$ & $1.25$ & $5.4\cdot10^4$ & $1.60 \pm 0.14$ & $5.02$ \\
\textbf{78} & $1.88 \pm 0.20$ & $0.94$ & $8.7\cdot10^4$ & $3.67 \pm 0.28$ & $1.05$ & $8.7\cdot10^4$ & $2.46 \pm 0.16$ & $5.2$ \\
\textbf{79} & $1.53 \pm 0.17$ & $1.32$ & $1.4\cdot10^5$ & $3.32 \pm 0.26$ & $0.96$ & $1.4\cdot10^5$ & $2.07 \pm 0.14$ & $5.8$ \\
\textbf{80} & $1.62 \pm 0.18$ & $1.13$ & $1.3\cdot10^5$ & $3.57 \pm 0.27$ & $0.77$ & $1.3\cdot10^5$ & $2.21 \pm 0.15$ & $6.11$ \\
\textbf{81} & $1.21 \pm 0.15$ & $1.22$ & $5.5\cdot10^4$ & $2.82 \pm 0.25$ & $1.29$ & $5.5\cdot10^4$ & $1.66 \pm 0.13$ & $5.5$ \\
\textbf{82} & $1.36 \pm 0.16$ & $1.07$ & $5.0\cdot10^4$ & $3.21 \pm 0.28$ & $1.04$ & $5.0\cdot10^4$ & $1.82 \pm 0.14$ & $5.79$ \\
\textbf{83} & $1.17 \pm 0.15$ & $1.08$ & $4.6\cdot10^4$ & $2.98 \pm 0.35$ & $0.91$ & $4.6\cdot10^4$ & $1.45 \pm 0.14$ & $4.71$ \\
\textbf{84} & $1.41 \pm 0.16$ & $0.96$ & $3.7\cdot10^4$ & $3.15 \pm 0.28$ & $0.89$ & $3.7\cdot10^4$ & $1.86 \pm 0.14$ & $5.4$ \\
\textbf{85} & $1.75 \pm 0.18$ & $1.12$ & $7.9\cdot10^4$ & $3.90 \pm 0.30$ & $1.17$ & $7.9\cdot10^4$ & $2.33 \pm 0.16$ & $6.07$ \\
\textbf{86} & $1.98 \pm 0.21$ & $0.95$ & $6.0\cdot10^4$ & $3.62 \pm 0.29$ & $0.79$ & $6.0\cdot10^4$ & $2.54 \pm 0.17$ & $4.63$ \\
\textbf{87} & $1.57 \pm 0.17$ & $0.84$ & $1.1\cdot10^5$ & $3.57 \pm 0.27$ & $1.03$ & $1.1\cdot10^5$ & $2.14 \pm 0.15$ & $6.16$ \\
\textbf{88} & $0.97 \pm 0.14$ & $1.47$ & $7.1\cdot10^5$ & $2.28 \pm 0.20$ & $0.91$ & $7.1\cdot10^5$ & $1.37 \pm 0.11$ & $5.31$ \\
\textbf{89} & $0.83 \pm 0.13$ & $1.11$ & $2.4\cdot10^6$ & $1.87 \pm 0.18$ & $1.21$ & $2.4\cdot10^6$ & $1.17 \pm 0.10$ & $4.62$ \\
\textbf{90} & $0.76 \pm 0.12$ & $0.88$ & $6.7\cdot10^5$ & $1.50 \pm 0.17$ & $1.1$ & $6.7\cdot10^5$ & $1.01 \pm 0.10$ & $3.6$ \\
\textbf{91} & $0.88 \pm 0.13$ & $1.17$ & $1.3\cdot10^6$ & $1.51 \pm 0.17$ & $1.39$ & $1.3\cdot10^6$ & $1.11 \pm 0.10$ & $3.01$ \\
\textbf{92} & $1.28 \pm 0.16$ & $1.01$ & $3.0\cdot10^4$ & - & - & $3.0\cdot10^4$ & $1.28 \pm 0.16$ & - \\
\textbf{93} & $1.47 \pm 0.17$ & $0.73$ & $2.1\cdot10^4$ & - & - & $2.1\cdot10^4$ & $1.47 \pm 0.17$ & - \\
\textbf{94} & $1.34 \pm 0.16$ & $1.02$ & $1.4\cdot10^4$ & - & - & $1.4\cdot10^4$ & $1.34 \pm 0.16$ & - \\
\textbf{95} & $1.78 \pm 0.20$ & $1.23$ & $3.7\cdot10^4$ & $4.09 \pm 0.35$ & $0.94$ & $3.7\cdot10^4$ & $2.36 \pm 0.18$ & $5.67$ \\
\textbf{96} & $1.79 \pm 0.20$ & $1.42$ & $1.6\cdot10^4$ & - & - & $1.6\cdot10^4$ & $1.79 \pm 0.20$ & - \\
\textbf{97} & $1.76 \pm 0.24$ & $0.98$ & $7.8\cdot10^4$ & $3.22 \pm 0.26$ & $1.56$ & $7.8\cdot10^4$ & $2.44 \pm 0.18$ & $4.1$ \\
\textbf{98} & $1.64 \pm 0.19$ & $0.96$ & $4.0\cdot10^4$ & $3.55 \pm 0.32$ & $1.25$ & $4.0\cdot10^4$ & $2.13 \pm 0.16$ & $5.2$ \\
\textbf{99} & $1.50 \pm 0.19$ & $1.68$ & $8.3\cdot10^3$ & - & - & $8.3\cdot10^3$ & $1.50 \pm 0.19$ & - \\
\textbf{100} & $1.47 \pm 0.18$ & $1.0$ & $4.3\cdot10^3$ & - & - & $4.3\cdot10^3$ & $1.47 \pm 0.18$ & - \\
\textbf{101} & - & - & $2.1\cdot10^3$ & - & - & $2.1\cdot10^3$ & - & - \\
\textbf{102} & - & - & $3.8\cdot10^2$ & - & - & $3.8\cdot10^2$ & - & - \\
\textbf{103} & $1.87 \pm 0.22$ & $1.05$ & $4.5\cdot10^3$ & - & - & $4.5\cdot10^3$ & $1.87 \pm 0.22$ & - \\
\textbf{104} & $1.78 \pm 0.21$ & $1.0$ & $1.3\cdot10^4$ & - & - & $1.3\cdot10^4$ & $1.78 \pm 0.21$ & - \\
\textbf{105} & - & - & $3.2\cdot10^3$ & - & - & $3.2\cdot10^3$ & - & - \\
\textbf{106} & $0.79 \pm 0.12$ & $1.33$ & $7.5\cdot10^5$ & $1.69 \pm 0.18$ & $0.97$ & $7.5\cdot10^5$ & $1.09 \pm 0.10$ & $4.16$ \\
\textbf{107} & $0.98 \pm 0.14$ & $0.65$ & $5.8\cdot10^3$ & - & - & $5.8\cdot10^3$ & $0.98 \pm 0.14$ & - \\
\textbf{108} & $0.82 \pm 0.13$ & $1.05$ & $7.2\cdot10^5$ & $1.74 \pm 0.18$ & $1.11$ & $7.2\cdot10^5$ & $1.13 \pm 0.10$ & $4.17$ \\
\textbf{109} & - & - & $2.9\cdot10^3$ & - & - & $2.9\cdot10^3$ & - & - \\
\textbf{110} & $0.76 \pm 0.12$ & $1.24$ & $3.5\cdot10^5$ & $1.57 \pm 0.17$ & $1.0$ & $3.5\cdot10^5$ & $1.03 \pm 0.10$ & $3.85$ \\
\textbf{111} & $0.83 \pm 0.13$ & $2.09$ & $1.8\cdot10^6$ & $1.53 \pm 0.17$ & $0.98$ & $1.8\cdot10^6$ & $1.08 \pm 0.10$ & $3.33$ \\

%% file: TrackerAuthorList_2024_Phase2_PixelModule_RadiationLength.tex
\setlength{\parindent}{0pt}
\setlength{\parskip}{6pt plus 2pt minus 1pt}
\setlength{\emergencystretch}{3em}  
\providecommand{\tightlist}{%
  \setlength{\itemsep}{0pt}\setlength{\parskip}{0pt}}
\setcounter{secnumdepth}{0}
\ifx\paragraph\undefined\else
\let\oldparagraph\paragraph
\renewcommand{\paragraph}[1]{\oldparagraph{#1}\mbox{}}
\fi
\ifx\subparagraph\undefined\else
\let\oldsubparagraph\subparagraph
\renewcommand{\subparagraph}[1]{\oldsubparagraph{#1}\mbox{}}
\fi

\section*{The Tracker Group of the CMS collaboration}
\addcontentsline{toc}{section}{The Tracker Group of the CMS collaboration}

\textcolor{black}{\textbf{Institut~f\"{u}r~Hochenergiephysik, Wien, Austria}\\*[0pt]
W.~Adam, T.~Bergauer, K.~Damanakis, M.~Dragicevic, R.~Fr\"{u}hwirth\cmsAuthorMark{1}, H.~Steininger}

\textcolor{black}{\textbf{Universiteit~Antwerpen, Antwerpen, Belgium}\\*[0pt]
W.~Beaumont, M.R.~Darwish\cmsAuthorMark{2,3}, T.~Janssen, P.~Van~Mechelen
% ...
}

\textcolor{black}{\textbf{Texas~A\&M~University, College~Station, USA}\\*[0pt]
R.~Eusebi}

\textcolor{black}{\textbf{Vanderbilt~University, Nashville, USA}\\*[0pt]
P.~D'Angelo, W.~Johns}

% ...
1: Also at Vienna University of Technology, Vienna, Austria \\
2: Also at Institute of Basic and Applied Sciences, Faculty of Engineering, Arab Academy for Science, Technology and Maritime Transport, Alexandria, Egypt \\
3: Now at Baylor University, Waco, USA\\
% ...
